%% file: main.tex
\documentclass[10pt,conference]{IEEEtran} 
\IEEEoverridecommandlockouts
\usepackage{cite}
\usepackage{amsmath,amssymb,amsfonts}
\usepackage{graphicx}
\usepackage{textcomp}
\usepackage{xcolor}
\usepackage{algorithm}
\usepackage{algorithmic}
\usepackage{arydshln}
\usepackage{amsmath}
\usepackage{multirow}
\usepackage{multicol}
\usepackage{subcaption}
\usepackage{enumitem}
\usepackage{booktabs}
\usepackage{xspace}
\usepackage{caption}
\usepackage{url}
\usepackage{bm}
\setlist[enumerate]{nolistsep}

\newcommand{\tool}{CREAM\xspace}

\def\BibTeX{{\rm B\kern-.05em{\sc i\kern-.025em b}\kern-.08em
    T\kern-.1667em\lower.7ex\hbox{E}\kern-.125emX}}

\begin{document}

\title{Two Sides of the Same Coin: Exploiting the Impact of Identifiers in Neural Code Comprehension}

\author{\IEEEauthorblockN{Shuzheng Gao$^{1}$, Cuiyun Gao$^{1\ast}$, Chaozheng Wang$^{1}$, Jun Sun$^{2}$, David Lo$^{2}$, Yue Yu$^{3}$}

\IEEEauthorblockA{$^1$ School of Computer Science and Technology, Harbin Institute of Technology, Shenzhen, China}



\IEEEauthorblockA{$^2$ Singapore Management University, Singapore}

\IEEEauthorblockA{$^3$ National University of Defense Technology}

\IEEEauthorblockA{szgao98@gmail.com,
gaocuiyun@hit.edu.cn,
wangchaozheng@stu.hit.edu.cn,
junsun@smu.edu.sg,
davidlo@smu.edu.sg,\\
yuyue@nudt.edu.cn
}

\thanks{$^{\ast}$ Corresponding author. The author is also affiliated with Peng Cheng Laboratory and Guangdong Provincial Key Laboratory of Novel Security Intelligence Technologies.}
}

\pagestyle{plain}

\maketitle

\begin{abstract}
Previous studies have demonstrated that neural code comprehension models are vulnerable 
to identifier naming. 
By renaming as few as one identifier in the source code, 
the models would 
output completely irrelevant results,  
indicating that identifiers can be misleading for model prediction. 
However, identifiers are not completely detrimental to code comprehension, since the semantics of identifier names can be related to the program semantics.
Well exploiting the two opposite impacts of identifiers is essential for enhancing the robustness and accuracy of neural code comprehension, and still remains under-explored. 
In this work, 
we propose to model the impact of identifiers from a novel causal perspective, and propose a counterfactual reasoning-based framework named \tool. 
\tool explicitly captures the misleading information of identifiers through multi-task learning in the training stage, and reduces the misleading impact
by counterfactual inference in the inference stage. We evaluate \tool on three popular neural code comprehension tasks, including function naming, defect detection and code classification. Experiment results show that \tool not only significantly outperforms baselines in terms of robustness (e.g., +37.9\% on the function naming task at F1 score), 
but also achieve improved results on the original datasets (e.g., +0.5\% on the function naming task at F1 score).



\end{abstract}


\input{section/intro}
\input{section/background}

\input{section/approach}
\input{section/setup}

\input{section/result}
\input{section/discussion}
\input{section/literature}
\section{Conclusion and future work}\label{sec:conclusion}
In this paper, we present \tool, a counterfactual reasoning-based framework to eliminate the misleading impact of identifiers 
in neural code comprehension. 
\tool captures the misleading information 
in the training stage through multi-task learning and reduces it by counterfactual inference in the inference stage. \tool is flexible and easy to be applied to various tasks and basic models. The evaluation on three  popular tasks demonstrates the effectiveness of \tool on original test sets and its robustness to identifier renaming.
In the future, we will explore to apply more causal inference techniques to solve the challenges in code intelligence tasks. 

\textbf{Data availability}: The implementation repository of this work is publicly available at \textit{{\url{https://github.com/ReliableCoding/CREAM}}}.


\bibliographystyle{IEEEtran}
\bibliography{sample.bib}

\end{document}

%% file: section/intro.tex
\section{Introduction}\label{sec:intro}
With the rapid development of artificial intelligence techniques, automated code comprehension has drawn more and more attention in the software engineering community~\cite{DBLP:conf/icse/ZhangWZ0WL19,DBLP:conf/kbse/LiuLZJ20,DBLP:conf/kbse/WeiLLXJ20,DBLP:journals/nn/GuLGWZXL21,DBLP:conf/iwpc/ShuaiX0Y0L20}. Recently, many neural code comprehension models have been proposed to learn the code semantics with deep neural networks (DNNs), and achieved state-of-the-art performance on various tasks, such as code summarization~\cite{DBLP:conf/acl/IyerKCZ16,DBLP:conf/kbse/WeiLLXJ20,DBLP:conf/iwpc/HuLXLJ18}, function naming~\cite{DBLP:conf/iclr/AlonBLY19,DBLP:conf/iclr/ZugnerKCLG21,DBLP:conf/icse/Li0N21}, and defect detection~\cite{DBLP:conf/nips/ZhouLSD019,DBLP:journals/tosem/ZouZXLJY21,IVDETECT,DBLP:conf/icse/NguyenNNLTP22}.

\begin{figure}[t]
    \centering
    \begin{subfigure}[b]{0.48\textwidth}
        \centering
        \includegraphics[width=0.8\textwidth]{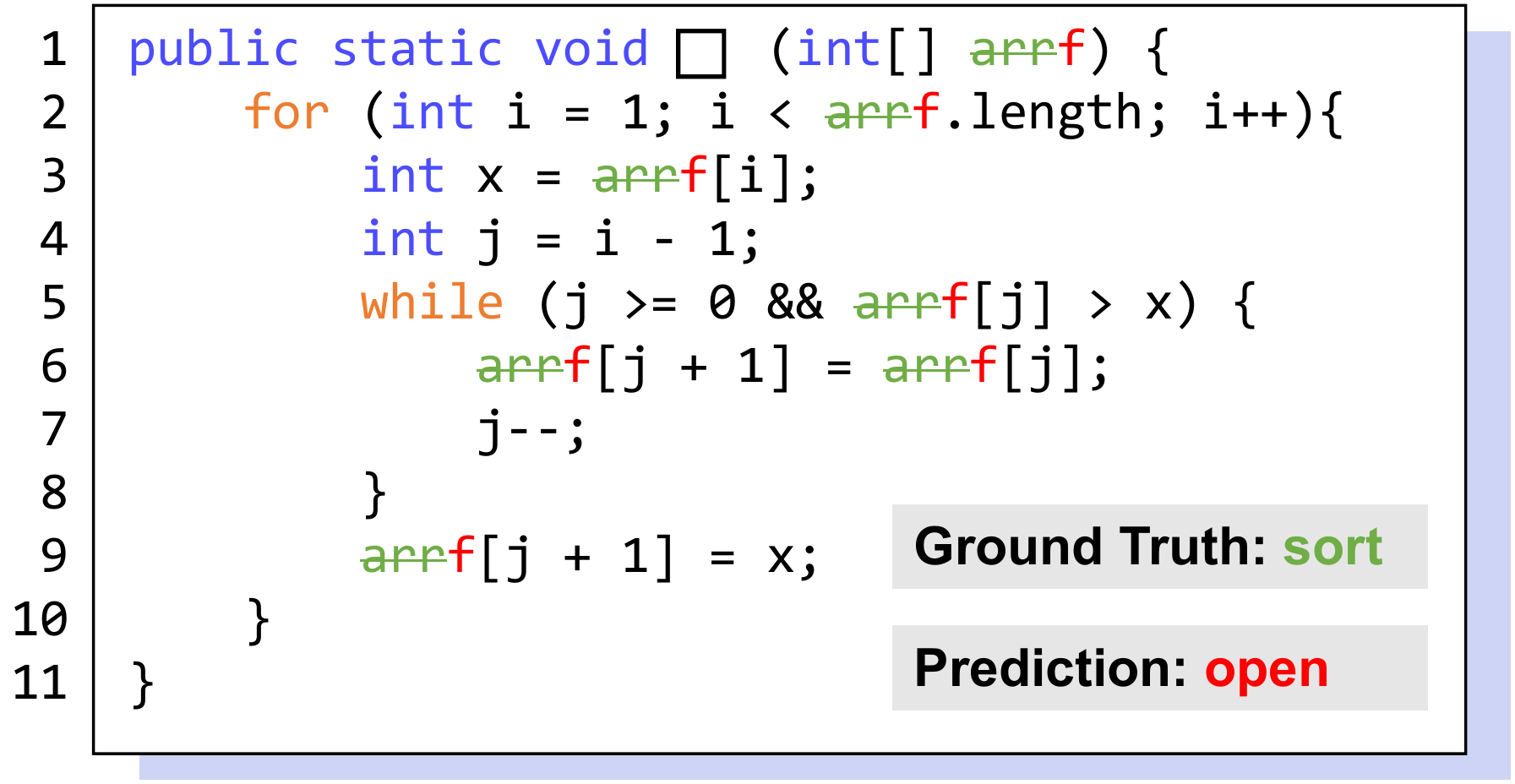}
        \caption{An example for illustrating that identifiers can be misleading for neural code comprehension models. By changing the identifier name ``\textit{arr}'' to ``\textit{f}'', the model outputs a wrong result.
        }
        \label{fig:example_1}
      \end{subfigure}
      \hfill
      \begin{subfigure}[b]{0.48\textwidth}
        \centering
        \includegraphics[width=0.8\textwidth]{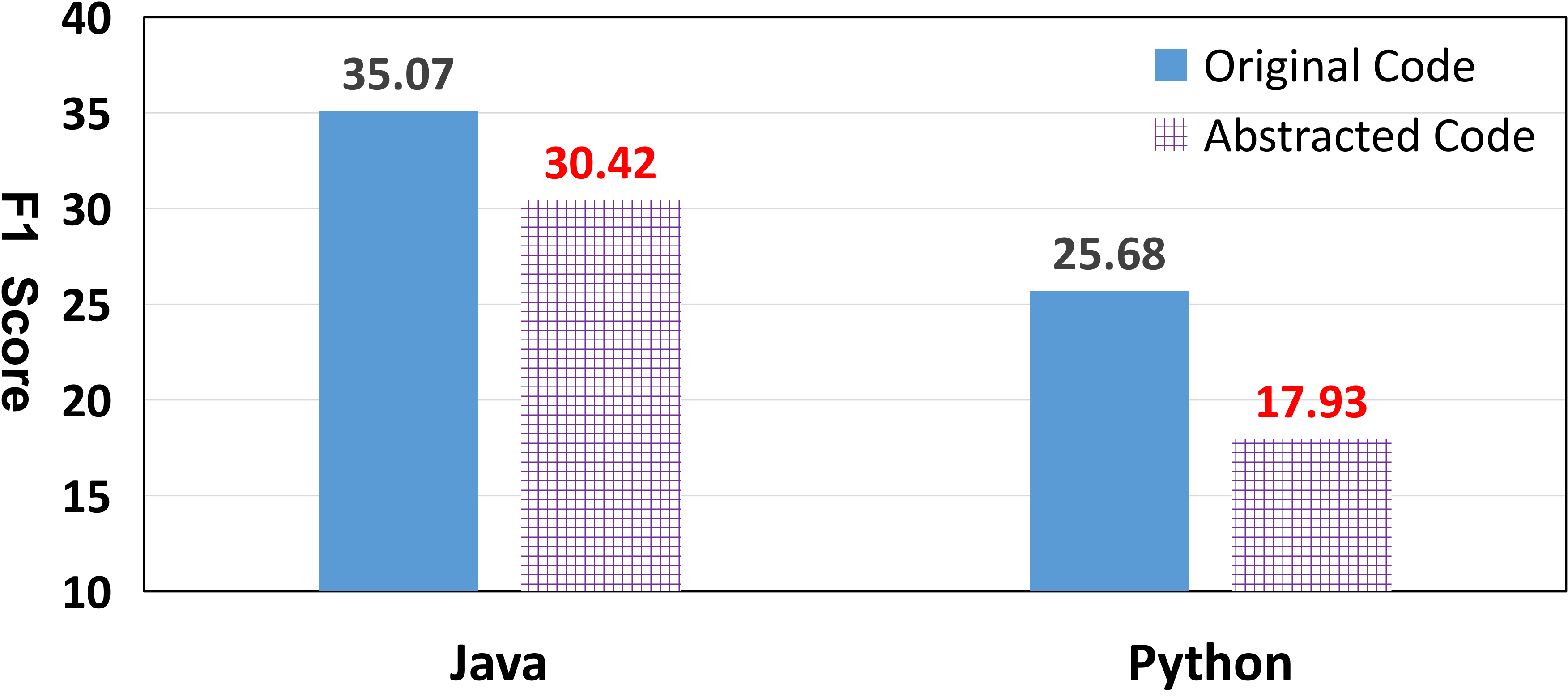}
        \caption{Results illustrating that code abstraction leads to performance drop on F1 score of the task.
        }
        \label{fig:example_2}
      \end{subfigure}
    \caption{Illustration of the opposite impact of identifiers on the function naming task.
    }
	\label{fig:motivated_example}
\end{figure}

Despite these success cases, 
recent studies~\cite{DBLP:journals/infsof/RabinBWYJA21,DBLP:conf/aaai/ZhangLLMLJ20} show that the  neural code comprehension models are sensitive to 
identifier naming. 
On the one hand, 
identifier names can be misleading for model prediction~\cite{DBLP:journals/pacmpl/Yefet0Y20,zhang2022towards,DBLP:journals/infsof/RabinBWYJA21,DBLP:journals/corr/abs-2002-03043}. 
For example,
as illustrated in Figure~\ref{fig:motivated_example} (a), given the original code snippet, the popular
function naming model NCS~\cite{DBLP:conf/acl/AhmadCRC20} 
correctly predicts the function name as ``\textit{sort}''. However, if 
the identifier name ``\textit{arr}'' is renamed to 
``\textit{f}'', the model outputs an irrelevant function name ``\textit{open}''. On the other hand, identifier names are not completely harmful to code comprehension and can provide rich information about the code semantics~\cite{DBLP:conf/sigsoft/ChirkovaT21}. As shown in~\cite{DBLP:conf/sigsoft/ChirkovaT21,DBLP:journals/pacmpl/Yefet0Y20}, 
discarding identifier semantics results 
in poor model performance. As illustrated in Figure~\ref{fig:motivated_example} (b), the code abstraction technique~\cite{DBLP:conf/sigsoft/ChirkovaT21,DBLP:conf/icse/MastropaoloSCNP21}, which renames all the identifiers with placeholders (e.g., ``VAR\_0''), 
leads to an obvious performance degradation on the prediction, showing 13.26\% and 30.18\% drop in terms of 
the F1 score for Java and Python, respectively. Therefore, the robustness issue 
caused by identifier names and their useful semantics are two sides of the same coin. Properly leveraging the useful information and alleviating the misleading impact of identifiers are 
essential for building an accurate and robust code comprehension model, which is a problem that 
remains under-explored.

\begin{figure}
    \centering
    \includegraphics[width=0.45\textwidth]{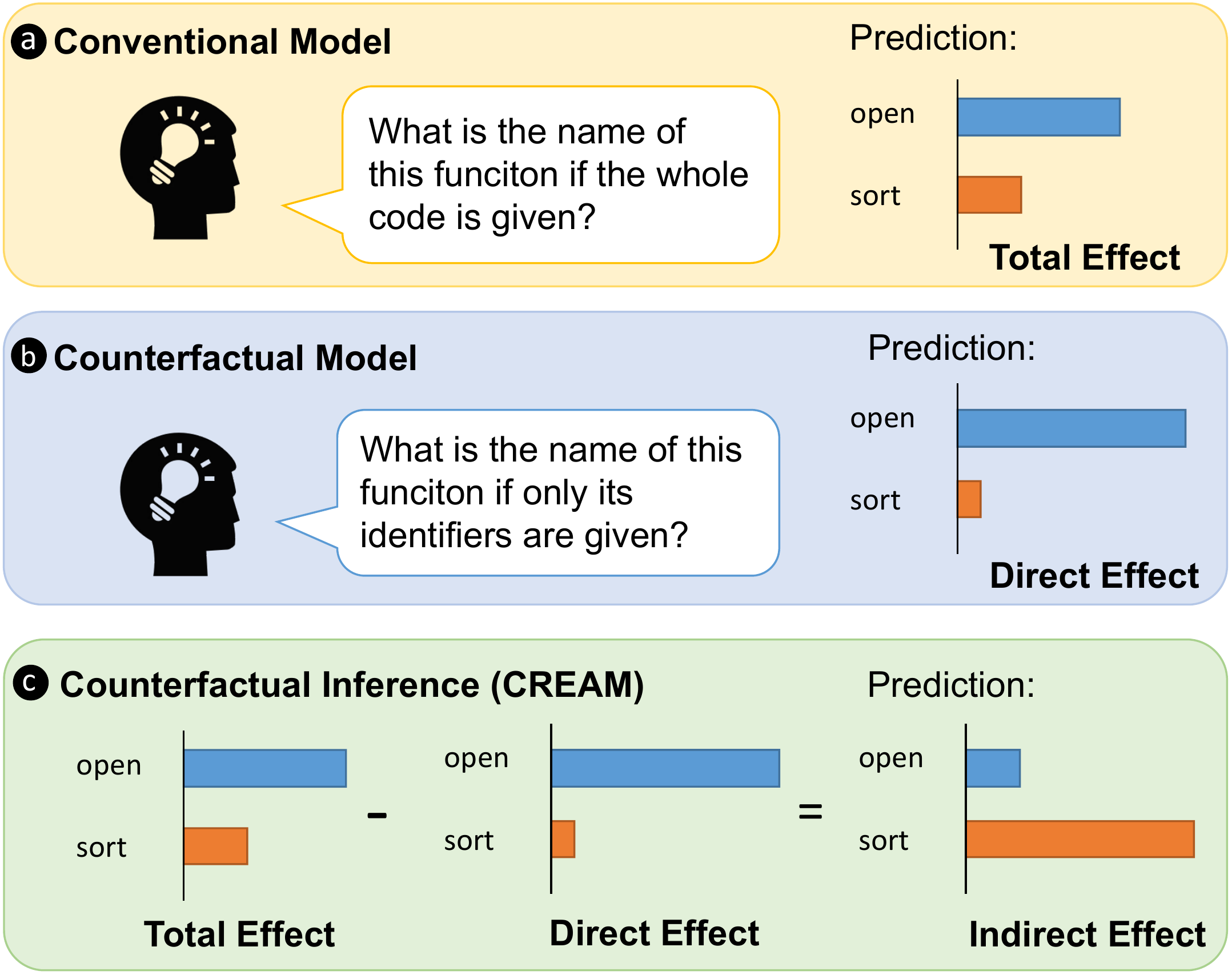}
    \caption{Our causal view on the impact of identifiers in neural code comprehension.
    The conventional model {(a)} and counterfactual model {(b)} estimate the result of corresponding question, respectively. Counterfactual inference (c) eliminates the \textit{direct effect} of identifiers, and 
    gets more accurate and robust prediction. 
    The prediction results are from the example in Figure~\ref{fig:motivated_example} (a).}
    \label{fig:overview}
\end{figure}

However, it is challenging to well balance the useful and misleading impact of identifiers on the model performance. 
The main challenge is that it is difficult to 
explicitly distinguish the two opposite impacts, the deep learning models are black-box at this stage~\cite{castelvecchi2016can}.
Besides, although producing more training instances might be helpful, the process would require non-trivial manual 
labeling efforts and extra training costs~\cite{schwartz2020green,zhu2009introduction}. In this work,  
we aim at exploiting the impact of identifiers without the cost of more training data.

Inspired by causal inference from the statistics field~\cite{pearl2009causality}, which has proven effective in analyzing opposite impacts in fields such as psychology~\cite{mackinnon2007mediation} and politics~\cite{keele2015statistics}, we propose to solve the problem from a causal view. Specifically,
the impact of the identifiers on model prediction can be divided into two separate causal effects, \textit{direct effect} and \textit{indirect effect}. By formulating the useful and misleading information as the \textit{direct effect} and \textit{indirect effect} respectively, 
we propose to preserve the useful information and mitigate the misleading impact of identifiers names by counterfactual inference~\cite{DBLP:conf/cvpr/NiuTZL0W21,DBLP:conf/kdd/WeiFCWYH21}, i.e. subtracting the direct identifier effect from the total effect. As illustrated in Figure~\ref{fig:overview} (a), conventional models perform prediction by estimating the \textit{total effect} of the input code snippet, and thereby are sensitive to identifier names. From the causal perspective, we first compute the \textit{direct effect} of identifiers by estimating ``\textit{what the name of this function would be if only its identifiers were given?}'', as shown in Figure~\ref{fig:overview} (b). The \textit{direct effect} is then removed through counterfactual inference for alleviating the misleading information of identifiers, as depicted in Figure~\ref{fig:overview} (c).


In this work, 
we propose \tool, a \textbf{C}ounterfactual \textbf{REA}soning-based fra\textbf{M}ework (\tool) to exploit the impact of identifiers
in neural code comprehension. Specifically, in the training stage, \tool distinguishes and captures \textit{direct effect} and \textit{indirect effect} by
multi-task learning. In the inference stage, \tool eliminates the misleading impact by reducing the direct identifier effect from the prediction.
We evaluate the performance of \tool on three popular code comprehension tasks, including function naming, software defect detection and code classification. For each task, we choose no fewer than 
three popular conventional models and enhance 
them with the proposed framework \tool. 
For evaluating the robustness improvement brought by removing the misleading information of identifiers,
we establish a transformed test set for each dataset by randomly substituting each identifier in the original test set with another one in the dataset. Experimental results demonstrate that the models equipped
with \tool not only significantly improve robustness on the transformed test sets but also achieve improved 
performance on the original test sets. 
Specifically, \tool improves the robustness of CodeBERT on the transformed test set by 37.9\%, 8.3\% and 1.9\% on function naming, defect detection, and code classification respectively; while on the original test set, \tool achieves an improvement of 0.5\%, 0.9\%, and 0.3\% on function naming, defect detection, and code classification respectively.
The main contributions of this paper are summarized as follows:
\begin{enumerate}
    \item To the best of our knowledge, we are the first to exploit the two-sided impact of identifiers in neural code comprehension from a causal view.
    \item We propose a novel counterfactual reasoning-based framework \tool for capturing the misleading information of identifiers via multi-task learning, and mitigating the misleading information via counterfactual inference.
    \item Extensive experiments show that the proposed framework \tool is more robust to identifier renaming than conventional models on various code comprehension tasks. The removal of misleading impact also improves the overall performance of \tool on the original datasets. 
\end{enumerate}

%% file: section/background.tex
\section{Preliminaries}\label{sec:back}
In this section, we review the key concepts we used in causal inference.
In the following, we use the capital letter (e.g., \textit{T}) and lowercase letter (e.g., \textit{t})  to denote a variable and a specific value respectively.

\begin{figure}
    \centering
    \includegraphics[width=0.45\textwidth]{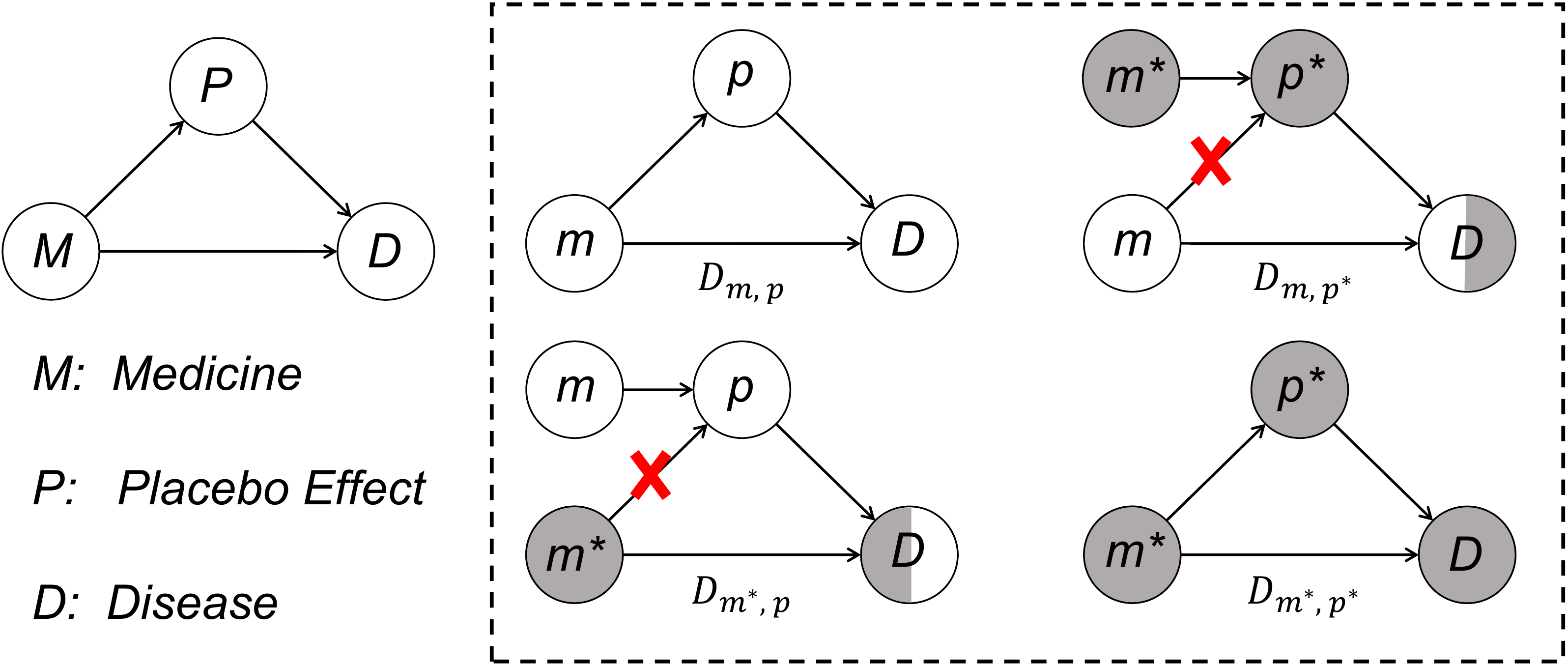}
    \caption{An example of SCM. White and gray nodes denote the variables are at the value of factual and counterfactual status respectively. Counterfactual notations $D_{m,p}$, $D_{m^*,p}$, $D_{m,p^*}$ and $D_{m^*,p^*}$ are illustrated in the four graphs on the right.
    }
    \label{fig:scm_example}
\end{figure}

\textbf{Structural Causal Model.}
The Structural Causal Model (SCM)~\cite{pearl2009causality,pearl2018book} reflects the causal relations between variables through a directed acyclic graph $G=\{V,E\}$, where $V$ denotes the set of variables and $E$ denotes the set of edges that describe the direct causal relationship between variables. These causal relationships can be further parameterized with \textit{structural equation}~\cite{pearl1998graphs}. Figure~\ref{fig:scm_example} is a simple example of SCM involving three variables, \textit{treatment variable} medicine ($M$), \textit{mediator variable} placebo effect 
($P$) and \textit{outcome variable} disease ($D$)~\cite{pearl2009causality}. Their causal relationships can be presented as follows:
\begin{equation}
    P_m = f_P(M=m),
\end{equation}
\begin{equation}
    D_{m,p} = f_D(M = m, P = p),
\end{equation}
where $f_P$ and $f_D$ denote the \textit{structural equation} of corresponding variable, respectively. In this case, the causal effect of medicine on disease exists in two paths. The first path \textit{M $\rightarrow$ D} denotes that medicine has a direct effect on disease through the biological mechanism. The other path \textit{M $\rightarrow$ P $\rightarrow$ D} shows that taking medicines can also alleviate the disease through the mediator, placebo effect. Thus when estimating to what extent the medicine affects the disease through the placebo effect, we need to exclude the bias caused by the biological mechanism of medicine, i.e., \textit{M $\rightarrow$ D}.

\textbf{Counterfactual Inference.} 
Counterfactual inference~\cite{pearl2009causality,pearl2018book} is used to estimate for the same individual what the outcome variable would be if the value of some variables were different from the value we observed in the reality. 
As shown in Figure~\ref{fig:scm_example}, counterfactual inference can answer the following question: \textit{whether the disease of the patient could be alleviated if he didn't receive the placebo effect but took medicines}. Specifically, it estimates the value of $D$ when $P$ receives $M = m$ through \textit{$M$ $\rightarrow$ $P$}, while $D$ receives $M = m^*$ through \textit{$M$ $\rightarrow$ $D$}. Here we use the asterisk notation to represent the situation where the value of the node is muted from the reality, e.g., $p^*$ denotes that the patient did not receive placebo effect.
This estimation can be achieved by using $do(P=p^*)$:
\begin{equation}
    D_{m,p^*} = f_D(M = m, do(P = p^*)),
\end{equation}
where $do(\cdot)$ operator denotes the intervention defined by SCM~\cite{glymour2016causal,pearl2009causality}. It forcibly substitutes $p = f_P(M=m)$ with $p^* = f_P(M=m^*)$ in the structural equation $f_D$. Note that $do(P = p^*)$ does not affect the ascendant variable of $P$, i.e., $M$ retains its value $m$ on the direct path \textit{M $\rightarrow$ D}. In clinical trials, it represents that the patient takes medicine without being informed. 

\textbf{Causal effects.} 
The causal effects measure to what extent value change of the treatment variable (e.g., the value of $M$ change from $m^*$ to $m$) affects the value of the outcome variable (e.g., $D$). For example in Figure~\ref{fig:scm_example}, the \textit{total effect} (TE)~\cite{pearl2009causality} of $M$ on $D$ is defined as:
\begin{equation}
    TE = D_{m, p} - D_{m^*, p^*},
\end{equation}
where $D_{m^*, p^*}$ denotes the situation that the patient neither took medicines nor received the placebo effect. We can see that TE calculates the causal effect of $M$ to $D$ from both the direct causal path \textit{M $\rightarrow$ D} and the indirect causal path \textit{M $\rightarrow$ P $\rightarrow$ D}. For a detailed analysis, existing work often decomposes TE into \textit{natural direct effect} (NDE) and \textit{total indirect effect} (TIE) through TE = NDE+TIE~\cite{vanderweele2013three,DBLP:conf/cvpr/NiuTZL0W21}. NDE represents the value change of the outcome variable when value change of the treatment variable only affects it through the direct path \textit{M $\rightarrow$ D}. Formally, NDE is defined as follows:
\begin{equation}
    NDE = D_{m, p^*} - D_{m^*, p^*},
\end{equation}
Accordingly, the TIE can be obtained by subtracting NDE from TE:
\begin{equation}
    TIE = TE - NDE = D_{m, p} - D_{m, p^*},
\end{equation}
TIE measures value change of the outcome variable when value change of the treatment variable only affects it through the indirect path \textit{M $\rightarrow$ P $\rightarrow$ D}. Equipped with the above causality concepts, we solve the problem of estimating the influence of placebo effect via calculating TIE of $M$ on $D$.

%% file: section/approach.tex
\section{METHODOLOGY}\label{sec:methoD}
In this section, we first present our causal view of the prediction process for neural code comprehension models, during which the impact of identifiers on the model prediction is
formulated with an SCM. 
Then we describe our proposed \tool framework for preserving the useful impact while eliminating the misleading impact of identifiers.

\subsection{A Causal View on Neural Code Comprehension}\label{subsec:scm}
As shown in Figure~\ref{fig:causal_graph} (a), we abstract the prediction process of neural code comprehension by defining
four variables: 1) \textbf{naming information \textit{T}}, which denotes the code tokens related to
identifier naming, i.e.,
user-defined identifiers;
2) \textbf{non-naming information \textit{F}}, which denotes the code properties irrelevant
to identifier naming, 
i.e., the code
tokens other than the identifiers;
3) \textbf{combined knowledge \textit{K}}, which serves as a mediator exploiting both the naming information and non-naming information for
model prediction;
4) \textbf{model prediction \textit{R}}, which denotes the prediction results of code comprehension tasks, e.g., classification scores for
code classification tasks. 

Based on the above variable definitions and SCM introduced in Section~\ref{sec:back}, we formulate the causal structure of conventional models, as illustrated in Figure~\ref{fig:causal_graph} (a). The causal relationships are shown as follows:

\begin{figure}[t]
    \centering
    \begin{subfigure}[b]{0.48\textwidth}
        \centering
        \includegraphics[width=0.8\textwidth]{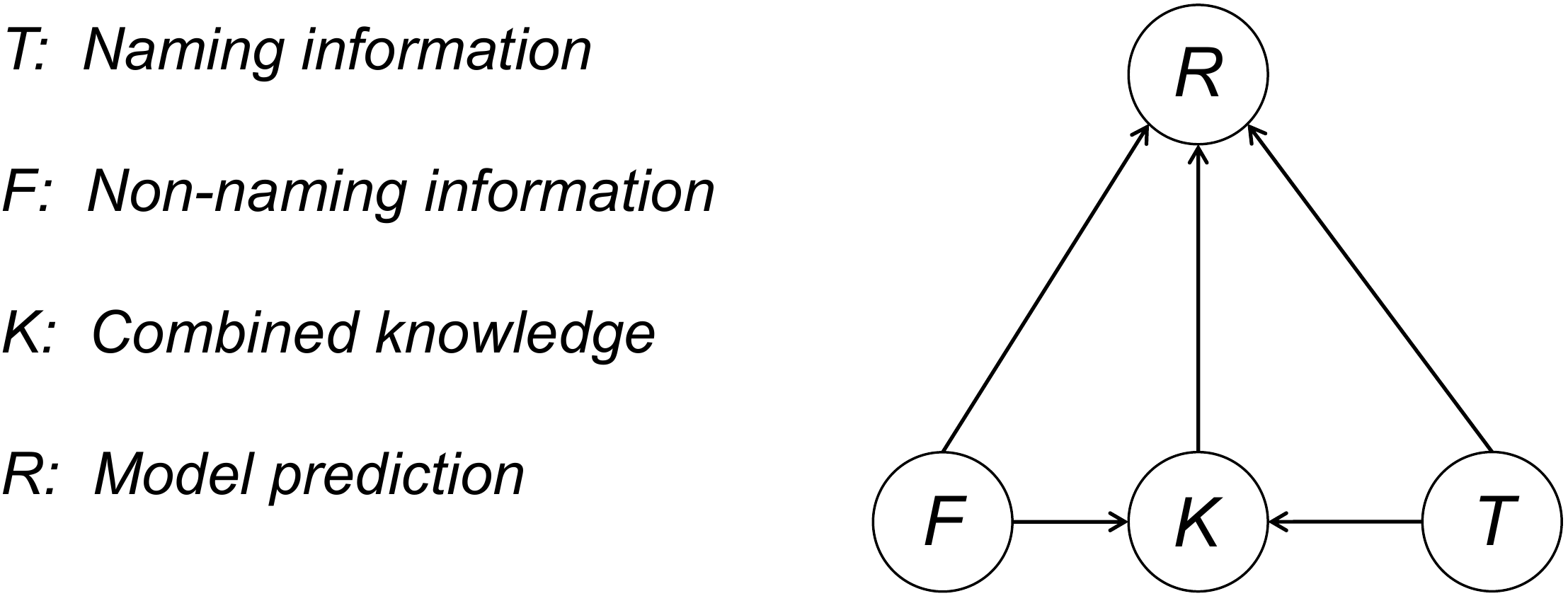}
        \caption{The SCM of conventional neural code comprehension model.}
        \label{tab:con_scm}
      \end{subfigure}
      \hfill
      \begin{subfigure}[b]{0.48\textwidth}
        \centering
        \includegraphics[width=0.8\textwidth]{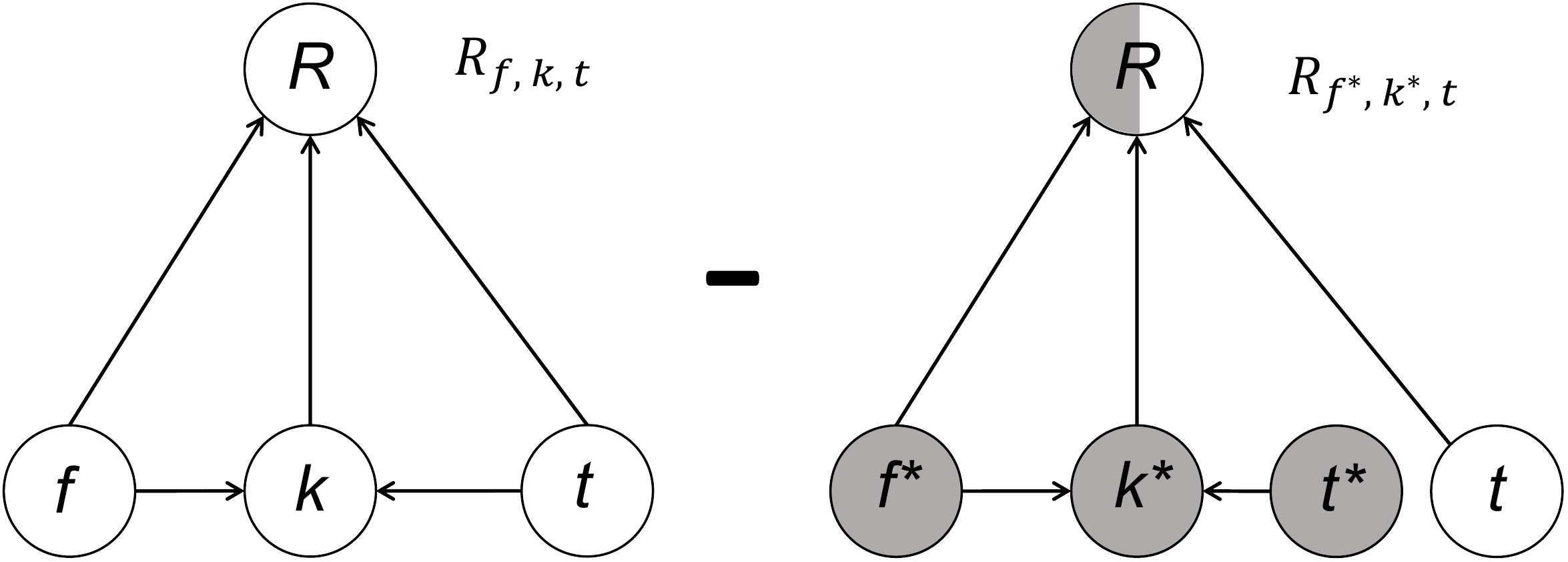}
        \caption{The SCM of our counterfactual reasoning-based neural code comprehension model.}
        \label{tab:cf_scm}
      \end{subfigure}
    \caption{Illustration for the SCM of conventional neural code comprehension model (a), and our counterfactual reasoning-based neural code comprehension model (b).}
	\label{fig:causal_graph}
\end{figure}

\begin{figure*}[ht]
    \centering
    \begin{subfigure}[b]{0.18\textwidth}
        \centering
        \includegraphics[width=1\textwidth]{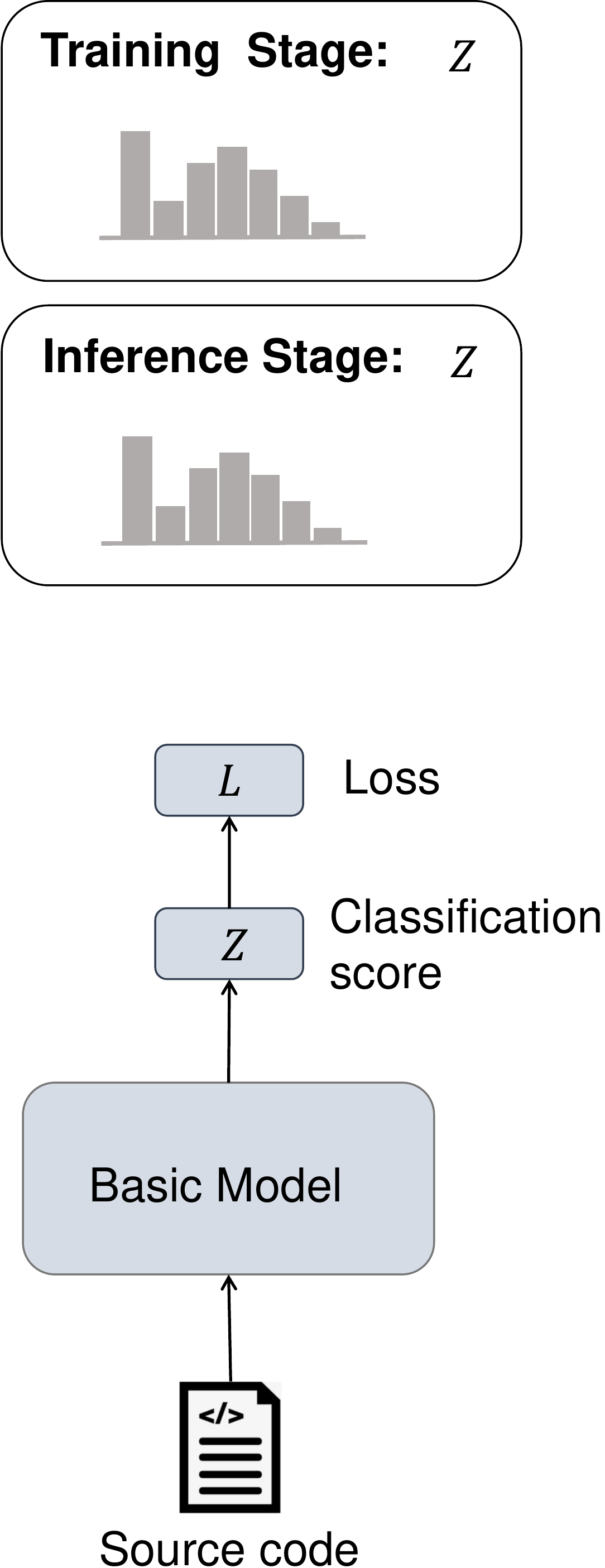}
        \caption{Conventional model.}
        \label{tab:con_framework}
      \end{subfigure}
      \hfill
      \begin{subfigure}[b]{0.80\textwidth}
        \centering
         \includegraphics[width=1\textwidth]{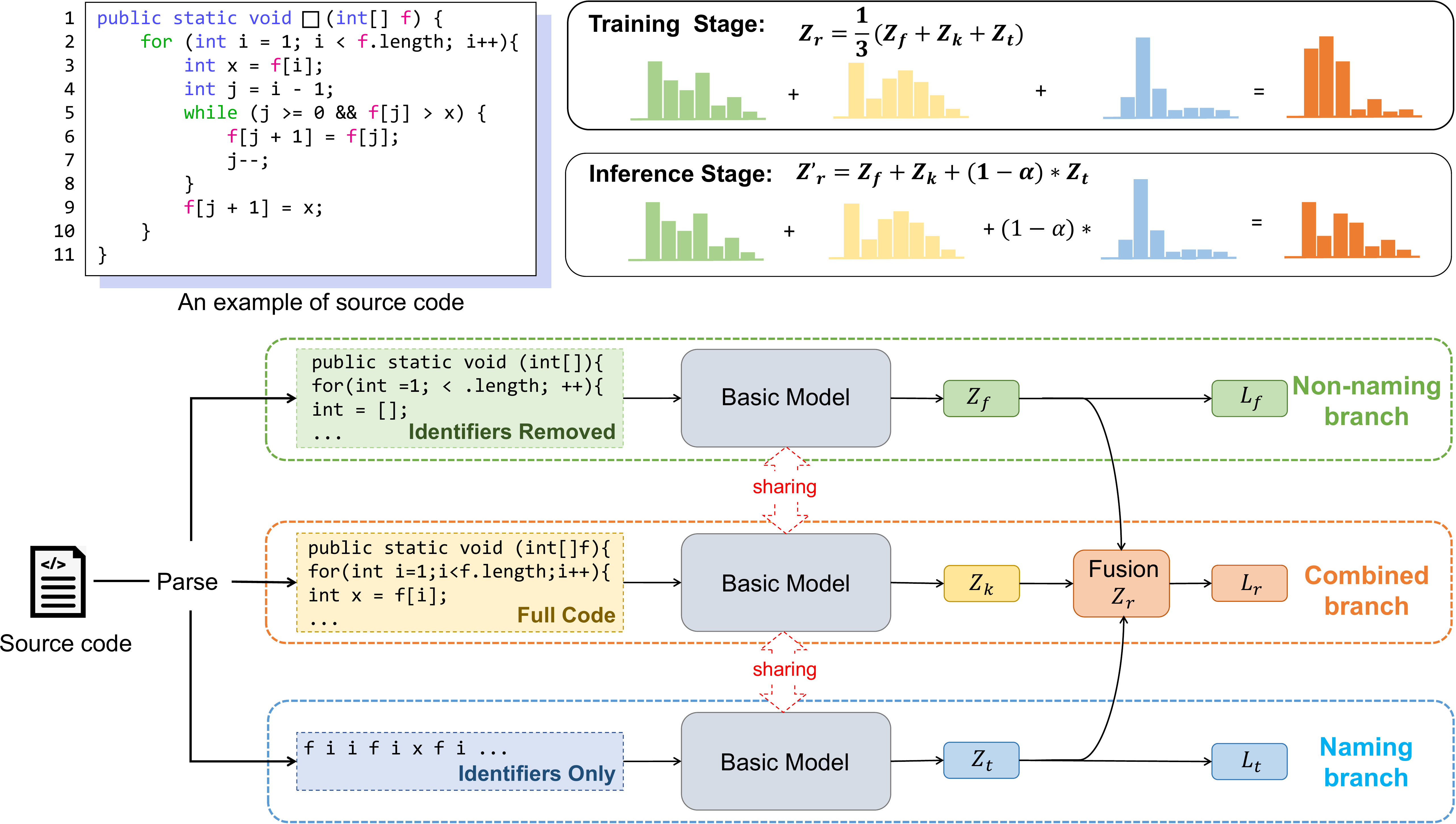}
        \caption{\tool.}
        \label{tab:cf_framework}
      \end{subfigure}
    \caption{The overall workflow of conventional neural code comprehension model (a) and \tool (b). The basic model in (a) and (b) denotes the
    conventional code comprehension models.} 
	\label{fig:framework}
\end{figure*}

\begin{enumerate}


\item \textbf{\textit{F $\rightarrow$ R}} represents the causal relationship from non-naming information $F$ to the model prediction $R$. For the example shown in Figure~\ref{fig:motivated_example} (a), the code structure which contains two loops corresponds to the variable $F$. Since the code structure is about data sorting-related swap operations, it is helpful for predicting the function name as ``\textit{sort}''.
We regard this path as a ``beneficial'' path, since
the prediction
based on 
only the non-naming information is 
robust to identifier renaming.

\item \textbf{\textit{T $\rightarrow$ R}}
represents the causal relationship from naming information $T$ to the model prediction $R$, 
For the example in Figure~\ref{fig:motivated_example} (a), the identifier ``\textit{f}'' which corresponds to variable $T$, misleads the model to predict ``\textit{open}'' as the function name. 
The possible reason of the wrong prediction is that the identifier ``\textit{f}'' commonly appears in the functions named ``\textit{open}'' in the training set.
For example, in a Java function named ``\textit{open}'', ``\textit{f}'' is usually used as an abbreviation of a file handling object, e.g., ``\textit{File f = new File()}''. The phenomenon is also called \textit{spurious correlation}~\cite{pearl2009causality} between identifiers and prediction results. Since the \textit{spurious correlation} will mislead models' understanding of the code semantics, we regard this direct causal effect as the misleading impact of identifiers.

\item 
\textbf{\textit{F}, \textit{T} \textit{$\rightarrow$ K $\rightarrow$ R}}
represents the process that models predict based on
the combined knowledge $K$ which exploits both the naming information $T$ and non-naming information $F$. 
We also regard this path as a ``beneficial'' path since it not only leverages the robust non-naming information \textit{F} and also the useful information in identifier names \textit{T}.
Although identifier
names are essentially irrelevant to
program behaviors, the semantics are helpful for accurate code comprehension~\cite{DBLP:conf/sigsoft/ChirkovaT21,DBLP:journals/pacmpl/Yefet0Y20}.
The path represents that we leverage the useful information of identifiers through the combined knowledge $K$ instead of completely discarding the identifiers.

\end{enumerate}

As shown in Figure~\ref{fig:causal_graph} (a), conventional neural code comprehension models predict through the \textit{total effect}, i.e., $R_{f,k,t}$ - $R_{f^*,k^*,t^*}$, which integrates the \textit{direct effect} from all the three paths. In this way, the inclusion of the ``bad'' path \textit{T $\rightarrow$ R} will inevitably introduce the misleading impact of identifiers to the prediction of conventional models. 
Thus, to improve the robustness and accuracy of the models, the \textit{direct effect} of $T$ $\rightarrow$ $R$ from the \textit{total effect} should be excluded during model prediction. 
To achieve the goal, we introduce the counterfactual reasoning-based neural code comprehension model 
{which} 
estimates the causal effect of $T$ and $F$ on the prediction $R$ with \textit{direct effect} $T$ $\rightarrow$ $R$ blocked, as shown in Figure~\ref{fig:causal_graph} (b).
We describe how we implement the idea and details of the proposed general framework \tool for eliminating the misleading impact of identifiers 
in the next section.

\subsection{Details of \tool}
In this section, we elaborate on the details of \tool. Figure~\ref{fig:framework} (a) and (b) illustrate the overall workflow of the conventional model and the proposed \tool, respectively. For conventional models, the training and inference stages are the same; while for \tool, the calculation of classification scores 
\footnote{It is also called logits in many machine learning papers~\cite{DBLP:journals/corr/HintonVD15,DBLP:conf/iclr/MenonJRJVK21}.} 
in the training and inference stage are different. Specifically, in the training stage, \tool captures each \textit{direct effect} in SCM through multi-task learning;
while in the inference stage, it eliminates the misleading impact of identifiers 
by counterfactual inference.

\subsubsection{Task formulation}

In this work, we broadly divide
neural code comprehension tasks into two paradigms,
i.e., classification-based and generation-based. Considering that the generation-based task can be viewed as a successive classification task, where the models output classification scores over the vocabulary at each time step, we unify the workflow of the neural code comprehension tasks as follows.
Assume that we have a source code database $X=\{x_1, x_2, ..., x_n\}$ and corresponding ground truth
$Y=\{y_1, y_2, ..., y_n\}$, where $n$ is the number of training data and $y_j$ is a class label or target sequence of the $j$-th training instance
for classification and generation task, respectively. In the training stage, our goal is to train a neural model $\mathcal{F}$ to minimize the prediction error on the training set. Formally, $\mathcal{F}$
can be formulated as:

\begin{equation}
    \bar{\mathcal{F}} = \mathop{\arg\min}_{\mathcal{F}} \sum_{(x,y) \in \{X,Y\}}^{} L(\mathcal{F}(x), y),
\end{equation}
where $L(\cdot)$ denotes the loss function such as cross entropy and $\mathcal{F}$ can be a large variety of existing neural models such as
Long Short-Term Memory (LSTM)~\cite{hochreiter1997long} and Transformer~\cite{vaswani2017attention}. In the inference stage, given a sample $x'$ in the test set, the model first calculates the classification score, i.e., $Z$ and $Z'_r$ for the conventional model and \tool, respectively (as shown in Figure~\ref{fig:framework}). Following the widely-used greedy search strategy, the class with the highest classification score is
selected as prediction result:
\begin{equation}
     y_r = \mathop{\arg\max}_c(z'),
\end{equation}
where $c$ is the set of candidate classes and $y_r$ is the prediction result.
\subsubsection{Framework Design}~\label{subsubsec:framework}
As shown in Figure~\ref{fig:framework} (b), our proposed \tool framework follows the SCM illustrated in Section~\ref{subsec:scm}.
Specifically, we distinguish the three causal paths including \textit{F $\rightarrow$ R}, \textit{K $\rightarrow$ R} and \textit{T $\rightarrow$ R} by designing three branches in \tool, i.e., non-naming branch, combined branch, and naming branch, respectively. To avoid increasing the model size, 
the parameters of the basic models for the three branches are shared. 
\tool first parses the input source code into three types of input, including non-naming input $f$, combined input $k$, and naming input $t$ for the three branches, respectively. \tool then estimates the direct causal effect of each path by calculating the classification score for each branch based on the basic model, formulated as:

\begin{equation}
Z_f= \mathcal{F}(f), Z_k= \mathcal{F}(k), Z_t= \mathcal{F}(t).
\end{equation}

To obtain the \textit{total effect} of the input code
on prediction, we combine the direct causal effect from three branches and fuse their {outputs} 
into {the final classification score} $Z_r$ 
which is corresponding to the variable $R$ in SCM. Here we compute $Z_r$ by
an averagely weighted method
\begin{equation}
    Z_r = \frac{1}{3}(Z_f+Z_k+Z_t),
\end{equation}
According to the definition of \textit{structural equation} in Section~\ref{sec:back}, the fusion method also indicates that
the \textit{structural equation} of the variable $R$ is parameterized as follows:
\begin{equation}\label{equ:fusion}
    R_{f,k,t} = \frac{1}{3}(\mathcal{F}(f)+\mathcal{F}(k)+\mathcal{F}(t)).
\end{equation}

Based on the above framework design of \tool, we introduce the computation process which includes two stages, i.e., training stage and inference stage. The details are illustrated in Algorithm~\ref{alg:framework}. 

\textbf{Multi-task Training:}\label{subsubsec:training}
In the training stage, \tool needs to realize multiple goals:
1) to accurately estimate the \textit{total effect} of source code on prediction results; and 2) to capture the \textit{direct effect} of each path and distinguish it
from \textit{total effect}. 
To achieve the goals, we adopt
the multi-task learning strategy~\cite{DBLP:conf/sigir/WangF0ZC21,DBLP:conf/kdd/WeiFCWYH21} for model training:
\begin{equation}
    L_f=L(Z_f,y), L_r=L(Z_r,y), L_t=L(Z_t,y),
\end{equation}
\begin{equation}
    L_{total} = L_f+L_r+L_t,
\end{equation}
Here, $L_r$ is used to train the model for accurately estimating
the \textit{total effect} (i.e., the first goal); while $L_f$ and $L_t$ are involved
to capture the \textit{direct effect} and distinguish it
from the \textit{total effect} (i.e., the second goal). Besides, to 
simultaneously capture the \textit{direct effect} and distinguish it from \textit{total effect}, we also adopt the deferred training strategy~\cite{DBLP:conf/nips/CaoWGAM19}. Specifically, we first train the three branches separately to well capture each \textit{direct effect}, and then fuse their classification scores after $I_{fusion}$ iterations
(Lines 4-7 of the multi-task training stage in Algorithm~\ref{alg:framework}).

\textbf{Counterfactual Inference:}\label{subsubsec:infer}
As illustrated in Section~\ref{subsec:scm}, 
the key to eliminate the misleading impact of identifiers 
is to remove the \textit{direct effect} \textit{T $\rightarrow$ R} from the \textit{total effect}. 
To this end, during the inference stage, \tool
first estimates the \textit{total effect} of the input code on model prediction $R$ as follows:
\begin{equation}
    TE = R_{f,k,t} - R_{f^*,k^*,t^*},
\end{equation}
where $f^*$, $k^*$ and $t^*$ denote the corresponding empty input.
Following~\cite{DBLP:conf/cvpr/NiuTZL0W21,DBLP:conf/kdd/WeiFCWYH21}, we
set the classification score to a uniform distribution (i.e., the classification scores for all classes are the same) if the input is empty, which is formulated as:
\begin{equation}
\mathcal{F}(f^*) = \mathcal{F}(k^*) = \mathcal{F}(t^*) = u
\end{equation}
where $u$ is the classification score under
uniform distribution. Then \tool estimates the misleading impact 
by calculating the \textit{natural direct effect} 
of identifier names
on model prediction, i.e., the \textit{direct effect} of $T = t$ on prediction $R$ under the situation $F = f^*$ and $K = k^*$:
\begin{equation}
    NDE = R_{f^*,k^*,t} - R_{f^*,k^*,t}.
\end{equation}
We finally eliminate the misleading impact by
subtracting the \textit{natural direct effect}
from the \textit{total effect}.
We propose to control the degree of elimination by involving a hyper-parameter $\alpha$, defined as: 
\begin{equation}\label{equ:debias}
    TE - \alpha * NDE \propto R_{f,k,t} - \alpha * R_{f^*,k^*,t},
\end{equation} 
where $\alpha$ ranges
from 0 to 1.
By omitting the constant, the final classification score $Z'_r$ is calculated
as follows:
\begin{equation}
\begin{aligned}
 Z'_r&=R_{f,k,t} - \alpha * R_{f^*,k^*,t^*}\\
    &=\frac{1}{3}(\mathcal{F}(f)+\mathcal{F}(k)+\mathcal{F}(t)) - \frac{\alpha}{3}(\mathcal{F}(f^*)+\mathcal{F}(k^*)+\mathcal{F}(t))\\
    &=\frac{1}{3}(\mathcal{F}(f)+\mathcal{F}(k)+\mathcal{F}(t)) - \frac{\alpha}{3}(u+u+\mathcal{F}(t))\\
    &\propto \mathcal{F}(f) + \mathcal{F}(k) + (1-\alpha) * \mathcal{F}(t)\\
    &=Z_f + Z_k + (1-\alpha) * Z_t.
\end{aligned}
\end{equation}


%% file: section/setup.tex
\section{Experimental setup}\label{sec:setup}
In this section, we detail the experimental settings for the three popular code comprehension
tasks including function naming, defect detection and code classification, which have been active areas of software engineering research for years.

\subsection{Evaluation Tasks}
\subsubsection{Function naming}
Function naming aims to automatically generate a meaningful and succinct name for a function. 
In software industry, it can help engineers correct the inconsistent method and API name for program readability and maintainability~\cite{DBLP:conf/icse/NguyenPLN20,DBLP:conf/icse/Li0N21,DBLP:journals/corr/abs-2201-10705}. In this work, we formulate it as a generation task with the greedy search strategy and cross-entropy loss function.
\subsubsection{Defect detection}
Given a code snippet, defect detection aims to 
identify 
whether a given code snippet is vulnerable,
which is crucial to defend a software system from cyberattack~\cite{DBLP:journals/tosem/ZouZXLJY21,DBLP:conf/nips/ZhouLSD019}. In the previous work, it is formulated as a binary classification task and generally uses the binary cross-entropy~\cite{bishop2006pattern}
as the loss function.
\subsubsection{Code classification}
Code classification is the task of classifying a code snippet by its functionality, which is helpful for program comprehension and maintenance~\cite{DBLP:conf/aaai/MouLZWJ16,DBLP:conf/icse/ZhangWZ0WL19}. It is formulated as a multi-class classification task and
utilizes cross-entropy as the loss function.

\begin{algorithm}[t]
\caption{Algorithm of \tool framework}
\begin{algorithmic}\label{alg:framework}
\REQUIRE training set $\{X_{train}, Y_{train}\}$, test set $\{X_{test}\}$, fusion iteration threshold $I_{fusion}$, total iteration $I$\\
\ENSURE neural model $\mathcal{F}$, prediction result $y_r$\\
{\bf Multi-task Training:}\\
\end{algorithmic}
\begin{algorithmic}[1]
\FOR{$i$ $\in$ $\{1,...,I\}$}
    \STATE Extract $f$, $t$, $k$ from $X_{train}$
    \STATE $Z_f$ = $\mathcal{F}$($f$), $Z_t$ = $\mathcal{F}$($t$), $Z_k$ = $\mathcal{F}$($k$)
    \IF{$i$<$I_{fusion}$}
        \STATE  $Z_r$ = $Z_k$
    \ELSE
        \STATE  $Z_r$ = $\frac{1}{3}$ ($Z_f$+$Z_k$+$Z_t$)
    \ENDIF
    \STATE Calculating $L_r$, $L_t$, $L_f$ with $Y_{train}$
    \STATE Update model $\mathcal{F}$ with $L_{total} = L_r$+$L_t$+$L_f$
\ENDFOR
\STATE \textbf{return} model $\mathcal{F}$
\end{algorithmic}
\begin{algorithmic}
\STATE 
{\bf Counterfactual Inference:}
\end{algorithmic}
\begin{algorithmic}[1]
\STATE Extract $f$, $t$, $k$ from $X_{test}$
\STATE $Z_f$ = $\mathcal{F}$($f$), $Z_t$ = $\mathcal{F}$($t$), $Z_k$ = $\mathcal{F}$($k$)
\STATE  $Z_r$ = $Z_f+Z_k+(1-\alpha) *Z_t$
\STATE  $y_r = \mathop{\arg\max}(Z_r)$
\STATE \textbf{return} Prediction result $y_r$
\end{algorithmic}
\end{algorithm}

\subsection{Baselines}
\subsubsection{Function naming}
{For function naming, we adopt three well-known code-to-text models for evaluation.}
\textbf{CodeNN}~\cite{DBLP:conf/acl/IyerKCZ16} is a classical sequence-to-sequence model which generates source code summaries with an LSTM network and attention mechanism.
\textbf{NCS}~\cite{DBLP:conf/acl/AhmadCRC20} is a recent state-of-art-model on code summarization. 
\textbf{CodeBERT}~\cite{DBLP:conf/emnlp/FengGTDFGS0LJZ20} is a widely-used pre-trained model for source code. We fine-tune CodeBERT with the pre-trained encoder with an additional decoder training from scratch.
\subsubsection{Defect detection}
For defect detection, we follow the popular benchmark CodeXGLUE~\cite{DBLP:journals/corr/abs-2102-04664} and adopt the following models.  \textbf{TextCNN} \cite{DBLP:conf/emnlp/Kim14} and \textbf{BiLSTM Att} \cite{hochreiter1997long} are two widely used methods for text classification in NLP. Here, \textbf{BiLSTM Att} is the combination of BiLSTM and attention mechanism~\cite{DBLP:conf/naacl/YangYDHSH16}. Many defect detection works~\cite{DBLP:conf/ndss/LiZXO0WDZ18,DBLP:conf/icmla/RussellKHLHOEM18} employ them for model construction.
\textbf{Devign}~\cite{DBLP:conf/nips/ZhouLSD019} is proposed to learn the various vulnerability characteristics with a composite code property graph and graph neural network. We also use \textbf{CodeBERT} as the basic model since it has shown promising results on defect detection~\cite{DBLP:journals/corr/abs-2102-04664}.
\subsubsection{Code classification}
We adopt three representative works in this field as the basic model for \tool. \textbf{TBCNN}~\cite{DBLP:conf/aaai/MouLZWJ16} is a classical code classification model which captures structural information of the Abstract Syntax Tree (AST) with a tree-based convolution neural network. \textbf{ASTNN}~\cite{DBLP:conf/nips/ZhouLSD019} learns the code representation by splitting the large AST into a sequence of small statement trees. We also involve the pre-trained model \textbf{CodeBERT} for evaluation.

\begin{table}[t]
 \centering
 \aboverulesep=0ex
\belowrulesep=0ex
\caption{Statistics of the benchmark datasets.}
\label{tab:dataset}
\scalebox{1.0}{
\begin{tabular}{lrrr}
\toprule
Datasets & \textbf{Train } & \textbf{Validation} & \textbf{Test}\\ 
\midrule
{\ CSN-Java } & 164,923 & 5,183 & 10,955\\
{\ CSN-Python } & 251,820 & 13,914 & 14,014\\
{\ CSN-Go } & 167,288 & 7,325 & 8,122 \\ 
{\ CSN-PHP } & 241,241 & 12,982 & 14,014 \\ 
{\ CSN-Ruby } & 24,927 & 1,400 & 1,261 \\ 
{\ CSN-JavaScript } & 38,499 & 2,745& 2,232 \\ 
\midrule
{\ Defect Detection} & 21,854 & 2,732 & 2,732\\
\midrule
{\ Code Classification} & 31,200 & 10,400 & 10,400\\ 
\bottomrule
\end{tabular}}
\end{table}

\begin{table*}[t]
\centering
\caption{Experimental results (F1 score) on function naming. Percentages listed within parantheses are computed improvement or reduction in F1 score as compared with the results of the corresponding basic model. ``*" denotes statistical significance in comparison to the baselines (i.e., two-sided $t$-test with $p$-value $<0.05$)}\label{tab:name_results}
\aboverulesep=0ex
\belowrulesep=0ex
\scalebox{0.97}{
\begin{tabular}{llllllll}
\toprule
\textbf{Approach} & \multicolumn{1}{c}{\textbf{Java}} & \multicolumn{1}{c}{\textbf{Python}} & \multicolumn{1}{c}{\textbf{JavaScript}} & \multicolumn{1}{c}{\textbf{PHP}} & \multicolumn{1}{c}{\textbf{Go}} & \multicolumn{1}{c}{\textbf{Ruby}} & \multicolumn{1}{c}{\textbf{Average}}\\
\midrule
& \multicolumn{7}{c}{\textbf{Transformed test set}}\\
\midrule
{  CodeNN} & 23.05 & 4.00 & 4.72 & 17.61 & 17.79 & 5.98 &12.19\\
{  +\tool}  & \textbf{28.35}($\uparrow$22.99\%)$^*$ & \textbf{5.44}($\uparrow$36.00\%)$^*$ & \textbf{6.00}($\uparrow$27.12\%)$^*$ & \textbf{22.44}($\uparrow$27.43\%)$^*$ & \textbf{24.91}($\uparrow$40.02\%)$^*$ & \textbf{7.25}($\uparrow$21.24\%)$^*$& \textbf{15.73}($\uparrow$29.04\%)\\
\hdashline
{ NCS} & 21.61 & 3.65 & 3.32 & 17.56 & 21.68 & 5.70 &12.25\\
{ +\tool} & \textbf{27.28}($\uparrow$26.24\%)$^*$ & \textbf{4.67}($\uparrow$27.95\%)$^*$ & \textbf{7.20}($\uparrow$116.87\%)$^*$ & \textbf{24.38}($\uparrow$38.84\%)$^*$ & \textbf{27.33}($\uparrow$26.06\%)$^*$ & \textbf{10.47}($\uparrow$83.68\%)$^*$ & \textbf{16.89}($\uparrow$37.88\%)\\
\hdashline
{ CodeBERT}  & 25.90 & 5.05 & 4.61 & 23.43 & 27.72  & 20.72 &17.91\\
{ +\tool}  & \textbf{36.72}($\uparrow$41.78\%)$^*$ & \textbf{7.29}($\uparrow$44.36\%)$^*$ & \textbf{10.54}($\uparrow$128.63\%)$^*$ & \textbf{35.80}($\uparrow$52.80\%)$^*$ & \textbf{38.62}($\uparrow$39.32\%)$^*$ & \textbf{27.49}($\uparrow$32.67\%)$^*$ & \textbf{26.08}($\uparrow$45.62\%)\\
\midrule
& \multicolumn{7}{c}{\textbf{Original test set}}\\
\midrule
{  CodeNN}  & 33.18 & 22.64 & \textbf{14.40}& 36.83 & 34.64 & 12.24 & 25.66\\
{  +\tool}  & \textbf{33.69}($\uparrow$1.54\%)$^*$ & \textbf{22.69}($\uparrow$0.22\%) & {14.16($\downarrow$1.67\%)} & \textbf{36.87}($\uparrow$0.11\%) & \textbf{34.79}($\uparrow$0.43\%) & \textbf{12.36}($\uparrow$0.98\%) &\textbf{25.76}($\uparrow$0.39\%)\\
\hdashline
{ NCS} & 35.07 & \textbf{25.68} & 16.31 & 40.19 & \textbf{40.52} & 12.93 & 28.45\\
{ +\tool} & \textbf{35.63}($\uparrow$1.59\%)$^*$ & {25.40($\downarrow$1.09\%)} & \textbf{16.46}($\uparrow$0.92\%) & \textbf{40.25}($\uparrow$0.04\%) & {40.21($\downarrow$0.77\%)} & \textbf{13.64}($\uparrow$5.5\%)$^*$ & \textbf{28.60}($\uparrow$0.53\%)\\
\hdashline
{ CodeBERT}  & 46.38 & 37.64 & \textbf{29.55} & 49.36 & 49.88 & \textbf{32.04} & \textbf{40.81}\\
{ +\tool}  & \textbf{47.04}($\uparrow$1.42\%)$^*$ & \textbf{37.66}($\uparrow$0.05\%) & {27.11($\downarrow$8.26\%)} & \textbf{50.35}($\uparrow$2.01\%)$^*$ & \textbf{50.28}($\uparrow$0.80\%)$^*$ & {30.49($\downarrow$4.84\%)} & {40.49($\downarrow$0.78\%)}\\
\bottomrule
\end{tabular}
}
\end{table*}

\begin{table}[t]
\centering
\caption{Experimental results (accuracy) on defect detection. ``*" denotes statistical significance in comparison to the baselines (i.e., two-sided $t$-test with $p$-value $<0.05$)}\label{tab:svd_results}
\aboverulesep=0ex
\belowrulesep=0ex
\scalebox{1}{
\begin{tabular}{lll}
\toprule
{\textbf{Approach}} & \textbf{Original test set} & \textbf{Transformed test set} \\
\midrule
{ TextCNN}  & 58.57 & 54.36\\
{ +\tool}  & \textbf{59.96}($\uparrow$2.37\%) & \textbf{55.78}($\uparrow$2.61\%)\\
\hdashline
{ BiLSTM Att}  & 62.08 & 54.90\\
{ +\tool}  & \textbf{62.34}($\uparrow$0.42\%) & \textbf{56.55}($\uparrow$3.01\%)\\
\hdashline
{ Devign}  & 55.77 & 52.03\\
{ +\tool}  & \textbf{56.74}($\uparrow$1.74\%) & \textbf{53.98}($\uparrow$3.75\%)\\
\hdashline
{ CodeBERT}  & 63.47 & 57.24\\
{ +\tool}  & \textbf{64.05}($\uparrow$0.91\%) & \textbf{62.01}($\uparrow$8.33\%)$^*$\\
\bottomrule
\end{tabular}
}
\end{table}

\begin{table}[t]
\centering
\caption{Experimental results (accuracy) on code classification. ``*" denotes statistical significance in comparison to the baselines (i.e., two-sided $t$-test with $p$-value $<0.05$)}\label{tab:cls_results}
\aboverulesep=0ex
\belowrulesep=0ex
\scalebox{1}{
\begin{tabular}{lll}
\toprule
{\textbf{Approach}} & \textbf{Original test set} & \textbf{Transformed test set} \\
\midrule
{ TBCNN}  & 96.76 & 68.45\\
{ +\tool}  & \textbf{97.00}($\uparrow$0.24\%) & \textbf{83.08}($\uparrow$21.37\%)$^*$\\
\hdashline
{ ASTNN}  & 98.04 & 89.43\\
{ +\tool}  & \textbf{98.18}($\uparrow$0.14\%) & \textbf{96.06}($\uparrow$7.41\%)$^*$\\
\hdashline
{ CodeBERT}  & 97.96 & 95.76\\
{ +\tool}  & \textbf{98.28}($\uparrow$0.33\%) & \textbf{97.56}($\uparrow$1.88\%)$^*$\\
\bottomrule
\end{tabular}
}
\end{table}

\subsection{Datasets and Metrics}

\subsubsection{Function naming}
For function naming, we use the widely used CodeSearchNet (CSN)~\cite{DBLP:journals/corr/abs-1909-09436} dataset which contains six programming languages including Java, Python, Go, PHP, JavaScript and Ruby. Specifically, we use the cleaned dataset which is pre-processed and open sourced in CodeBERT~\cite{DBLP:conf/emnlp/FengGTDFGS0LJZ20}. For JavaScript, we further filter the samples without a function name.
To measure the similarity between generated function names and the reference names, we employ the standard metrics including Precision, Recall and F1.

\subsubsection{Defect detection}
We use the defect detection dataset
released by Devign~\cite{DBLP:conf/nips/ZhouLSD019}. The dataset
contains 27,318 C code snippets collected from the QEMU and FFmpeg projects. As for the dataset split, we use the benchmark open sourced by  CodeXGLUE~\cite{DBLP:journals/corr/abs-2102-04664}, in which the dataset is split into training set, validation set and test set in a proportion of 8:1:1.
Following~\cite{DBLP:conf/nips/ZhouLSD019,DBLP:conf/emnlp/0034WJH21}, we use accuracy
as the evaluation metric.

\subsubsection{Code classification}
For code classification, we use the POJ dataset~\cite{DBLP:conf/aaai/MouLZWJ16} which contains 52,000 code snippets of C language with 104 classes. It is collected from Online Judge (OJ) and code snippets in the same class are used to solve the same programming problem. We follow ASTNN~\cite{DBLP:conf/nips/ZhouLSD019} to split the dataset into training set, validation set and test set in a proportion of 3:1:1. We follow previous work~\cite{DBLP:conf/aaai/MouLZWJ16,DBLP:conf/icse/ZhangWZ0WL19} in this field and use accuracy
as the evaluation metric.

We list the statistics of the benchmark datasets 
in Table~\ref{tab:dataset}. When evaluating the robustness, we follow previous work~\cite{DBLP:conf/sigir/BuiYJ21,DBLP:journals/corr/abs-2111-10793,DBLP:conf/issta/ZengTZLZZ22} and validate whether the model can output the same results under semantic-preserving code transformations, i.e., identifier renaming. Therefore 
we also create a
transformed test set for each dataset. Specifically, following the procedure in previous work~\cite{DBLP:conf/sigir/BuiYJ21,DBLP:journals/corr/abs-2111-10793,DBLP:conf/issta/ZengTZLZZ22}, we randomly substitute the identifier names in the test set with another identifier name that appeared in the dataset.

\subsection{Implementation Details}
During the experiment, we reproduce each model either directly using the code released by the author or strictly following the steps described in their paper\footnote{For the baseline Devign in the defect detection, we reproduce the method based on the re-implementation code in~\cite{reveal} due to the lack of original code.
}. For a fair comparison, we make sure that the hyperparameters such as training epochs and learning rate for models with and without \tool are exactly the same. 
The value of $\alpha$ in Equ. (\ref{equ:debias}) is set as 0.4, 0.5, 0.6, 0.7 or 0.8 for different datasets. The $I_{fusion}$ is set as
10\% of total training iterations. We will discuss how we select parameters for each dataset in Section~\ref{subsec:param}.

When applying \tool to each baseline, we parse the source code into three sequences (Figure~\ref{fig:framework}) for the models that treat source code as plain text (e.g., NCS and CodeBERT) for the function naming task. For models that treat the source code as a tree~\cite{DBLP:conf/icse/ZhangWZ0WL19} or graph~\cite{DBLP:conf/nips/ZhouLSD019}, we divide the tree or graph into two parts which
contain the nodes with and without the naming information, respectively.
To extract the identifiers in source code, we first parse the code
into AST with tree-sitter\footnote{{\url{https://github.com/tree-sitter/tree-sitter}}},
and filter the leaf node according to its type and the type of its parent. 
For example, we extract the identifiers for C/C++ by selecting the leaf node whose type is ``\textit{identifier}'' and parent's type is not ``\textit{call\_expression}''. 

All the experiments are conducted on a server with 4 Nvidia Tesla V100 GPUs and 32 GB graphic memory. We run each baseline and \tool three times and report the best results.

%% file: section/result.tex
\section{Experimental Results}\label{sec:result}

In this section, we evaluate the performance of \tool by answering the following research questions:

\begin{enumerate}[label=\bfseries RQ\arabic*:,leftmargin=.5in]
    \item Does \tool improve the robustness of existing models?
    \item Is \tool beneficial for improving the
    accuracy of existing models?
    \item What is the impact of multi-task learning and counterfactual inference on the performance of \tool?
    \item How do different parameter settings affect the performance of \tool?
\end{enumerate}

\begin{table*}[t]
\centering
\caption{Ablation study. Best and second best results are marked in bold and underline respectively.}\label{tab:ablation}
\aboverulesep=0ex
\belowrulesep=0ex
\scalebox{1}{
\begin{tabular}{lcccccc}
\toprule
\multirow{2}{*}{\textbf{Approach}} & \multicolumn{2}{c}{\textbf{Function Naming (F1)}} & \multicolumn{2}{c}{\textbf{Defect Detection (accuracy)}} & \multicolumn{2}{c}{\textbf{Code Classification (accuracy)}}\\
\cmidrule{2-7} 
& \multicolumn{1}{c}{Original} & \multicolumn{1}{c}{Transformed} & Original & Transformed & Original & Transformed\\
\midrule
 { CodeBERT}  & 49.36 & 23.43 & 63.47 & 57.24 & 97.96 & 95.76\\
 {+\tool}  & \textbf{50.35} & \underline{35.80} & \underline{64.05} & \textbf{62.01} & \underline{98.28} & \underline{97.56}\\
 \midrule
 { -w/o $L_f$}  & 49.67 & 32.80 & \textbf{64.20} & 58.38 & 98.12 & 89.90\\
 { -w/o $L_t$}  & 49.79& \textbf{36.68}  & 63.65 & \underline{61.42} & 98.16 & \textbf{98.02}\\
 { -w/o $L_t$ and $L_f$}  & 48.69 & 34.52 & 63.69 & 59.08 & 97.74 & 96.81\\
\hdashline
 { -w/o Counterfactual Inference}  & \underline{50.16} & 31.26 & 63.87 & 61.31 & \textbf{98.36} & 97.18\\
\bottomrule
\end{tabular}
}
\end{table*}

\subsection{RQ1: Evaluation on the Robustness of \tool}
We evaluate the robustness of \tool on the transformed datasets for the three tasks, with results illustrated in Table~\ref{tab:name_results}-\ref{tab:cls_results}, respectively. Due to the page limit, we only present the F1 scores of different models for the function naming task in Table~\ref{tab:name_results}. 
The scores for the precision and recall metrics are presented on the GitHub repository.\footnote{{\url{https://github.com/ReliableCoding/CREAM}}}. 
Based on the results, we have the following observations:


\textbf{Conventional models are not robust to identifier naming.}
By comparing the results
of conventional models on the original test set and transformed test set, we find that identifier renaming leads to significant performance degradation on the three tasks.
For example, the average performance of the models for function naming, defect detection, and code classification drop by 55.4\%, 8.9\% and 12.3\%, respectively. The results indicate that conventional neural code comprehension models are easily misled by identifier names. 

\textbf{\tool consistently and significantly improves the model robustness.} As can be seen in Table~\ref{tab:name_results}-\ref{tab:cls_results}, \tool consistently outperforms the conventional models on all the task datasets, demonstrating its capability and generalizability in improving the robustness of neural code comprehension models. Besides, the improvement over the conventional models is substantial, for example, 
for function naming \tool improves the performance of CodeNN, NCS and CodeEBRT on the transformed test set by 29.0\%, 37.9\%, and 45.6\%, respectively; for defect detection and code classification \tool also improves the performance of CodeBERT on the transformed test set by 8.3\% and 1.9\%, respectively.
The results suggest that
\tool can eliminate the misleading impact of identifiers, 
 enabling the models reliable to identifier renaming.


\textbf{The robustness improvement on smaller datasets are more obvious.} As shown in Table~\ref{tab:name_results}, by analyzing the results on the datasets with different sizes for the function naming task, we find that \tool achieves higher improvement on the smaller datasets.
For example, \tool boosts NCS by 116.9\% and 83.7\% on JavaScript and Ruby, respectively. This may be attributed to that neural models are prone to overfitting on datasets with small sizes~\cite{DBLP:journals/csur/WangYKN20,DBLP:conf/iclr/ZhangBHRV17}, thus the direct misleading information is prominent. 
Our proposed \tool can render the models less affected 
by identifier names especially on small datasets.

\subsection{RQ2: Performance Evaluation}
In this section, we evaluate the accuracy of \tool on the code comprehension tasks. From Table~\ref{tab:name_results}-\ref{tab:cls_results}, we observe that \tool improves the performance of conventional models in most cases.



\textbf{Function Naming.}
As shown in Table~\ref{tab:name_results}, we find that \tool
improves the performance of each basic model in a vast majority of cases. Specifically, the average improvement of \tool over CodeNN and NCS is 0.4\% and 0.5\%, respectively, regarding the F1 score. Although the averaged F1 score for CodeBERT+\tool drops slightly,
the performance on most
languages still increases.
The results show the effectiveness of \tool on function naming.

\textbf{Defect Detection.} 
As shown in Table~\ref{tab:svd_results}, we can observe that \tool
improves the accuracy of all the basic models with an average improvement of 1.4\%. Specifically, CodeBERT+\tool and Devign+\tool outperform their corresponding baselines by 0.9\% and 1.7\%, respectively, which indicates that \tool can facilitate conventional models to capture the patterns of
vulnerable code snippets.

\textbf{Code Classification.}
As shown in Table~\ref{tab:cls_results}, we can observe a consistent improvement of \tool on different basic models. For example, although the performance of ASTNN and CodeBERT are strong enough, i.e., achieving 98.04\% and 97.96\% accuracy, respectively, \tool can further boost them by 0.1\% and 0.3\%, respectively. This indicates that \tool is also effective to
comprehend the code functionality.  

\begin{figure*}[t]
     \centering
     \begin{subfigure}[h]{0.235\textwidth}
        \centering
    	\includegraphics[width=1 \textwidth]{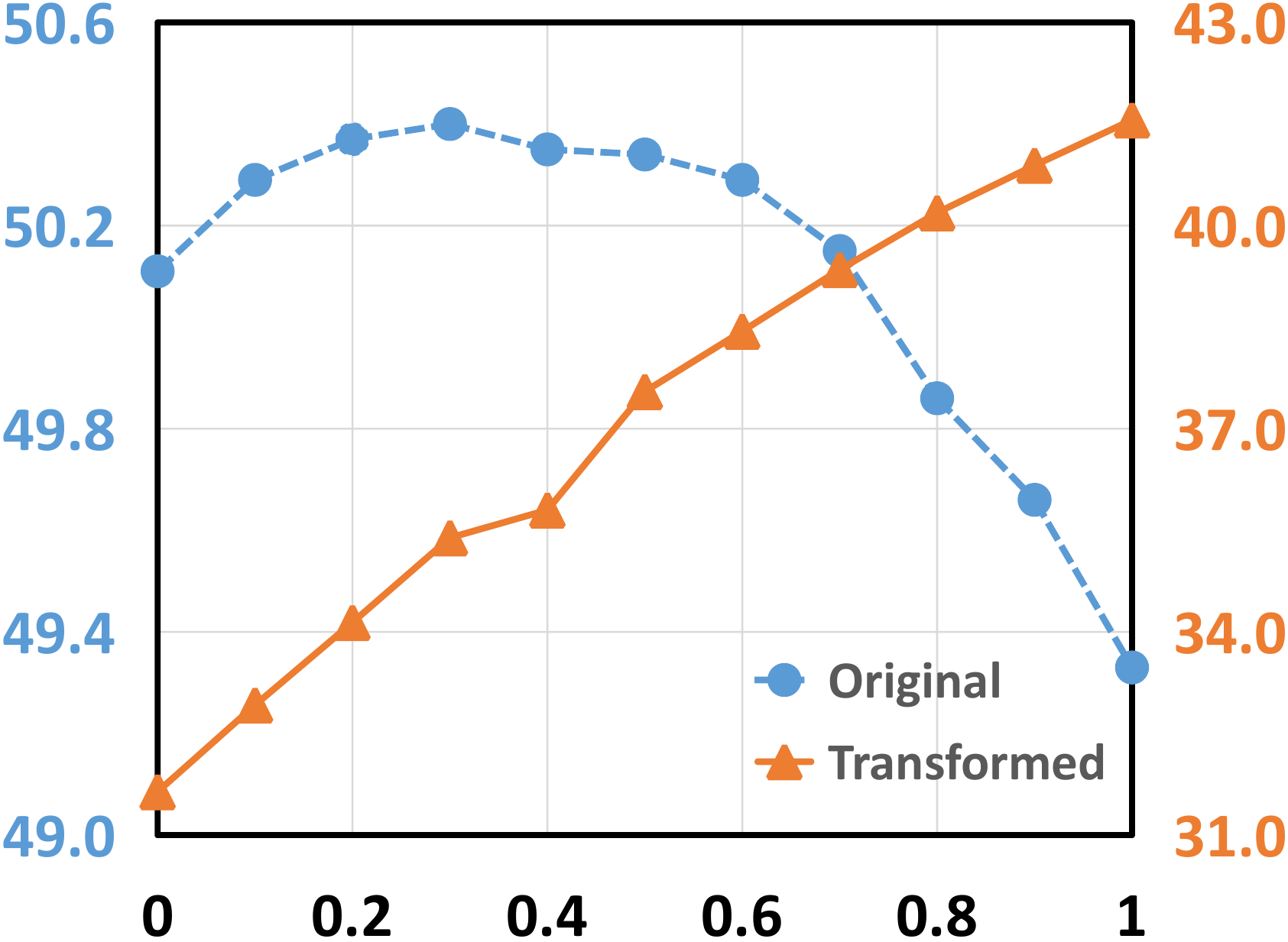}
    	\caption{\small $\alpha$ on PHP function naming.}
        \end{subfigure}
        \hfill
        \begin{subfigure}[h]{0.235\textwidth}
        \vspace*{0.15cm}
        \centering
    	\includegraphics[width=1 \textwidth]{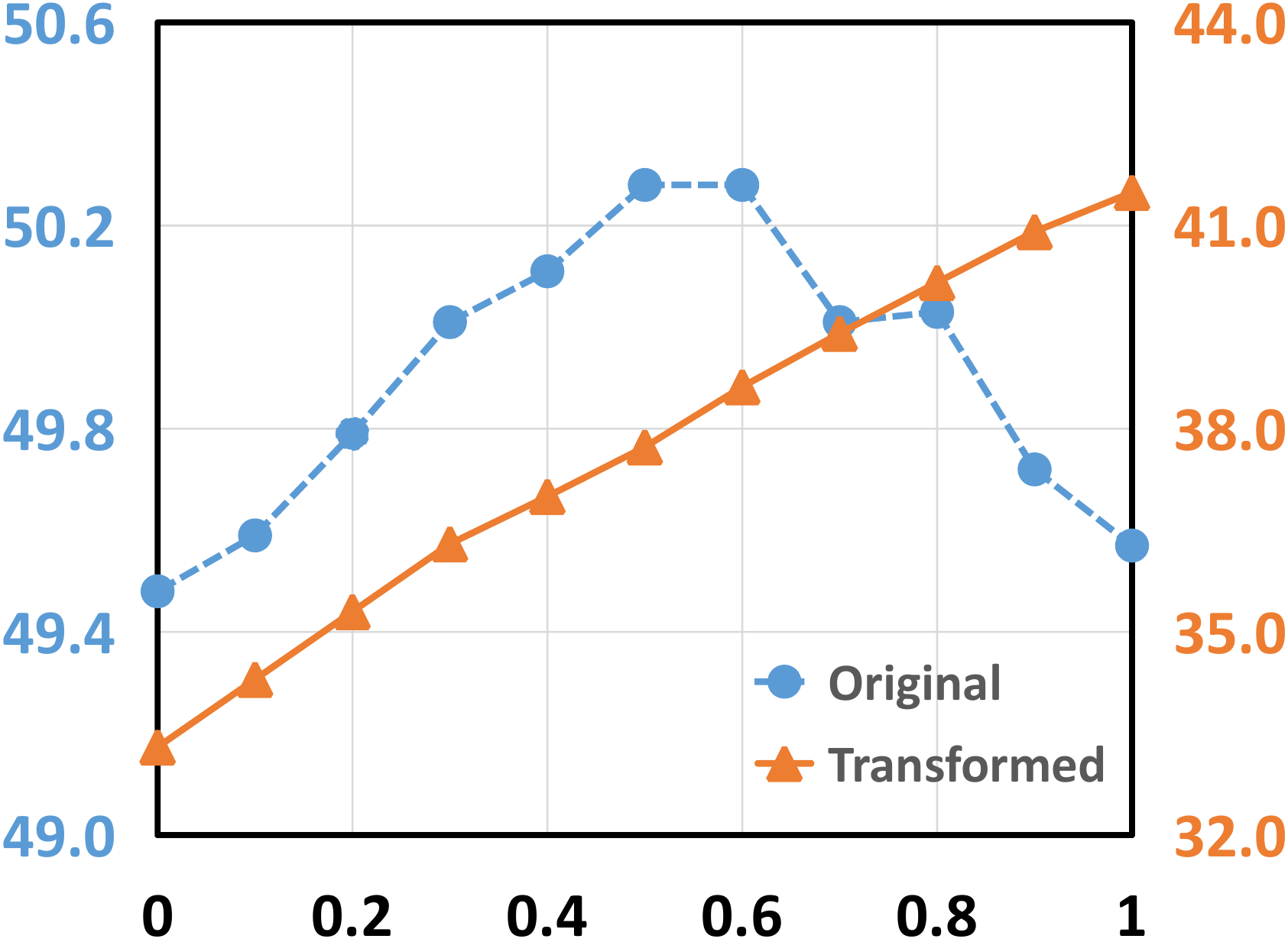}
    	\caption{\small $\alpha$ on GO function naming.}
        \end{subfigure}
        \hfill
        \begin{subfigure}[h]{0.235\textwidth}
        \vspace*{0.15cm}
        \centering
        \includegraphics[width=1 \textwidth]{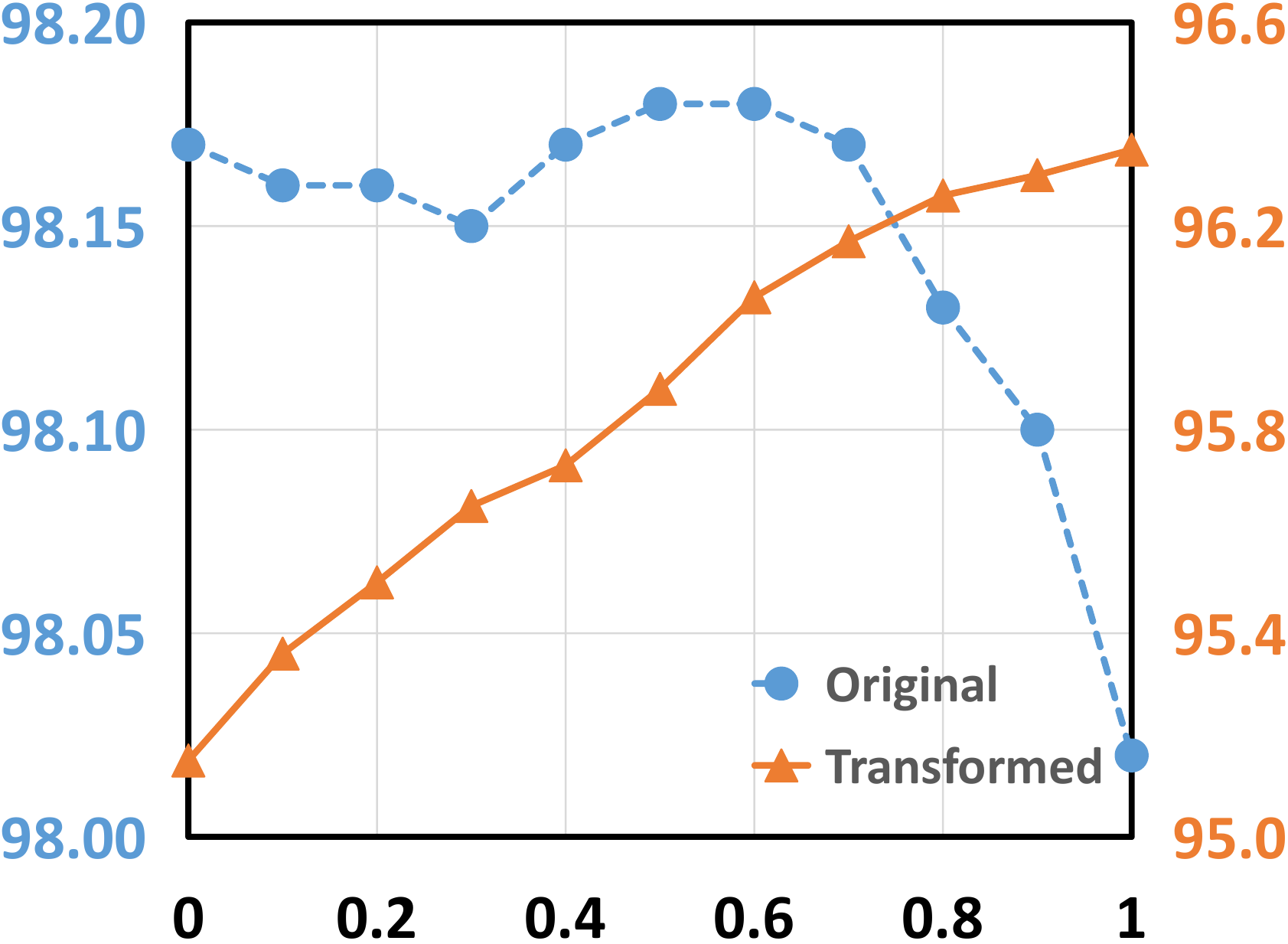}
        \caption{\small $\alpha$ on defect detection.}
        \end{subfigure}
        \hfill
        \begin{subfigure}[h]{0.235\textwidth}
        \vspace*{0.15cm}
        \centering
        \includegraphics[width=1 \textwidth]{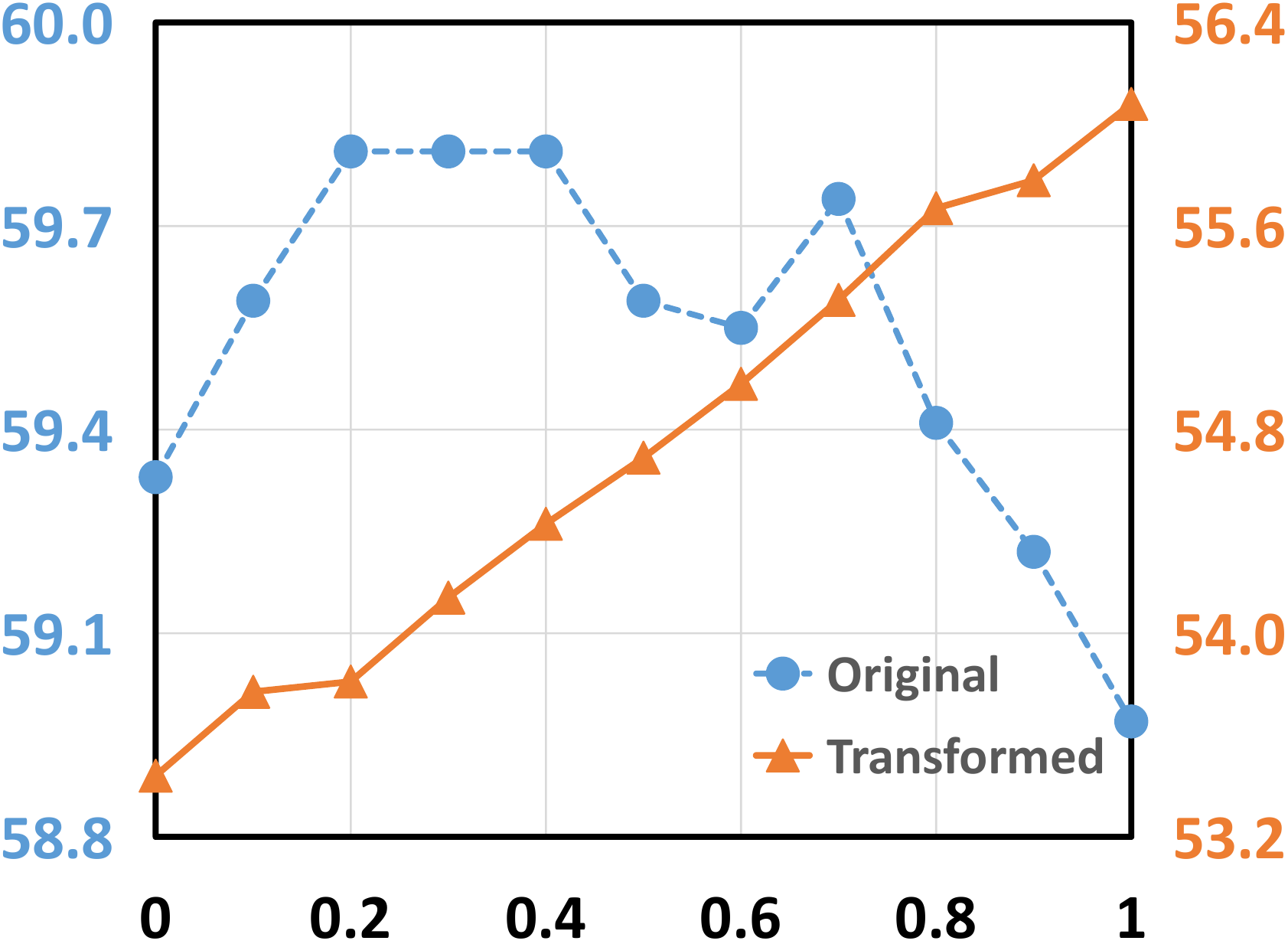}
        \caption{\small $\alpha$ on code classification.}
        \end{subfigure}
        \hfill
        \begin{subfigure}[h]{0.235\textwidth}
        \centering
        \includegraphics[width=1 \textwidth]{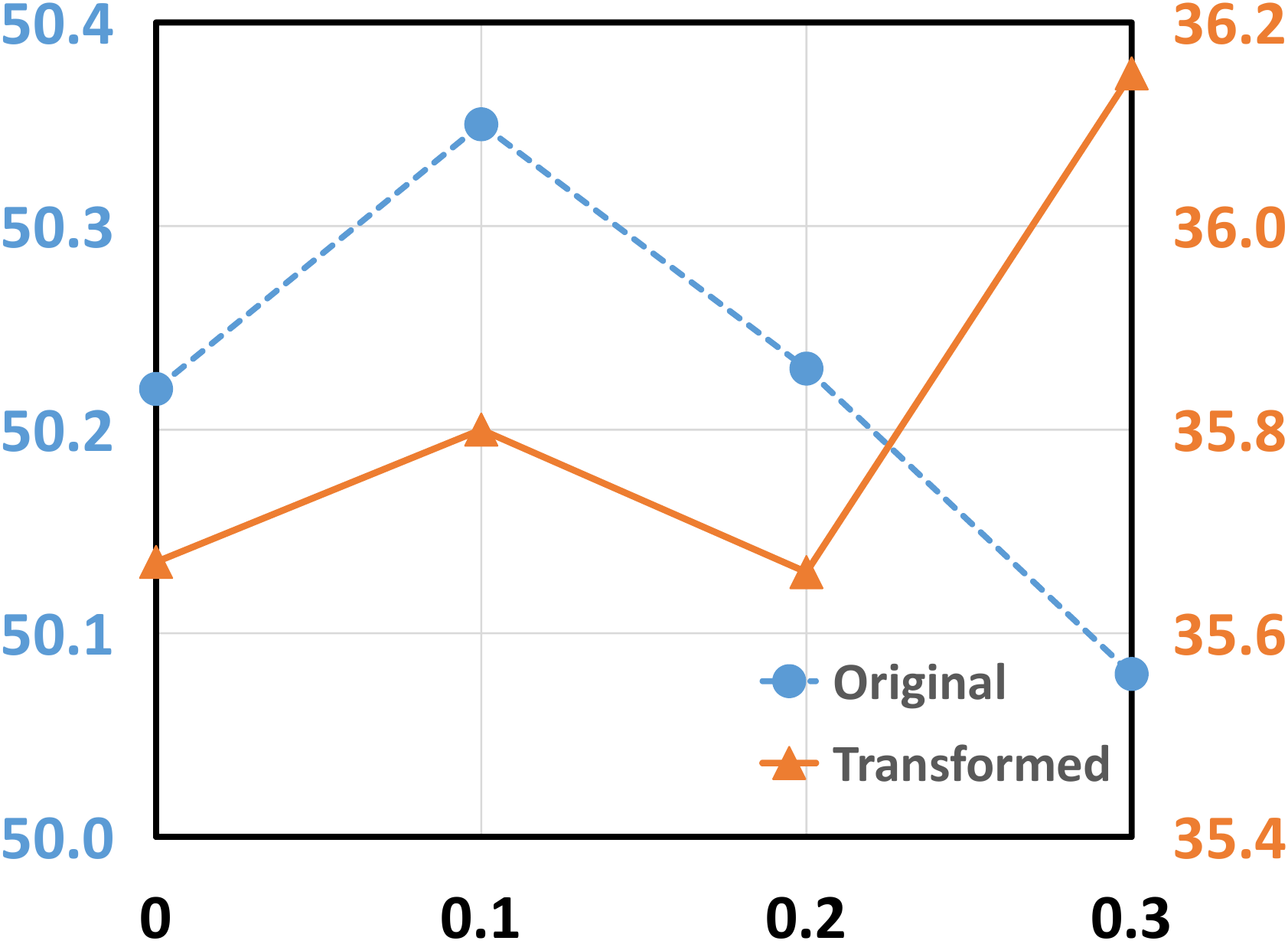}
        \caption{\small $I_{f}$ on PHP function naming.}
        \end{subfigure}
        \hfill
        \begin{subfigure}[h]{0.235\textwidth}
        \centering
        \includegraphics[width=1 \textwidth]{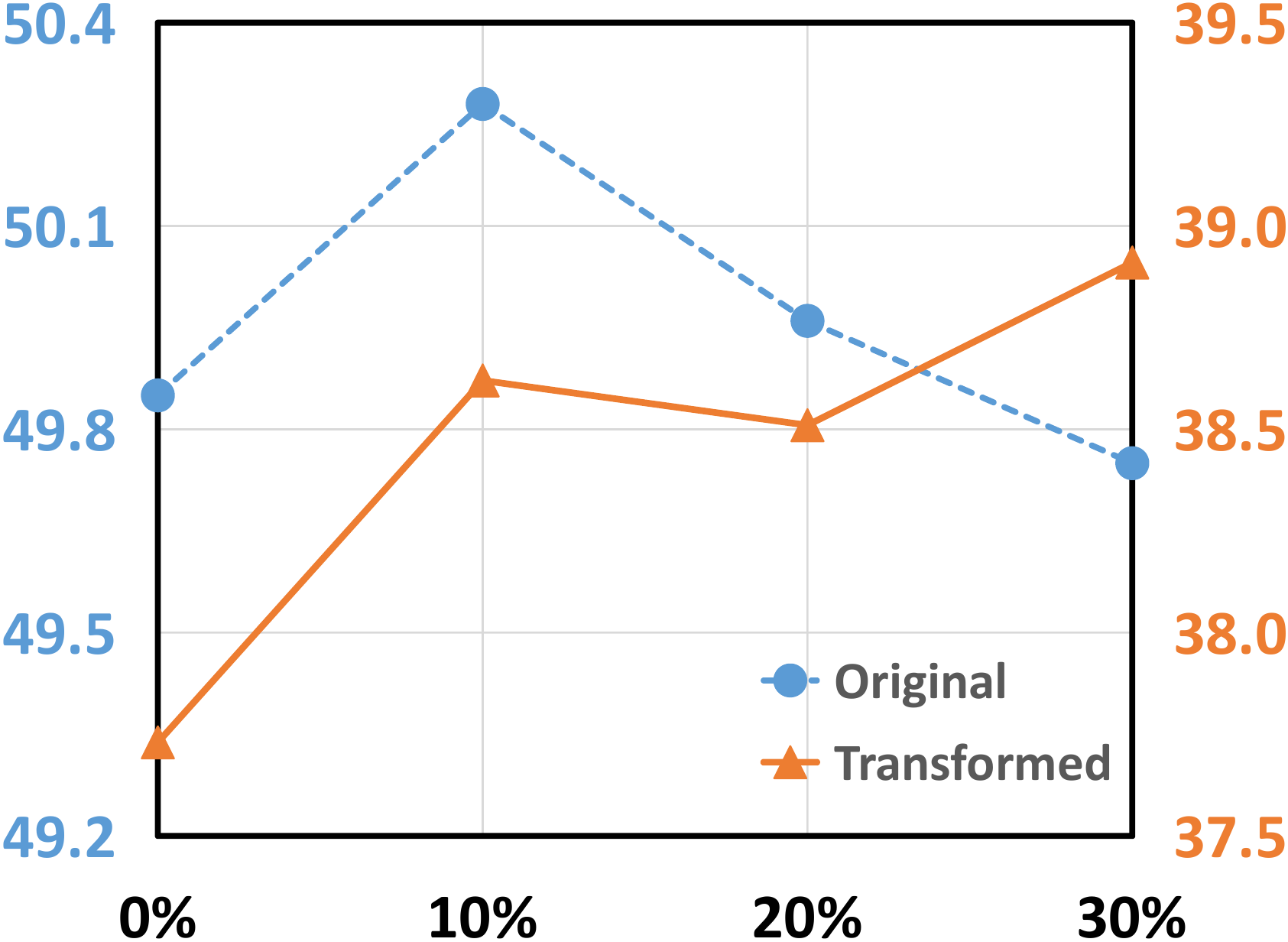}
        \caption{\small $I_{f}$ on GO function naming.}
        \end{subfigure}
        \hfill
        \begin{subfigure}[h]{0.235\textwidth}
        \vspace*{0.15cm}
        \centering
        \includegraphics[width=1 \textwidth]{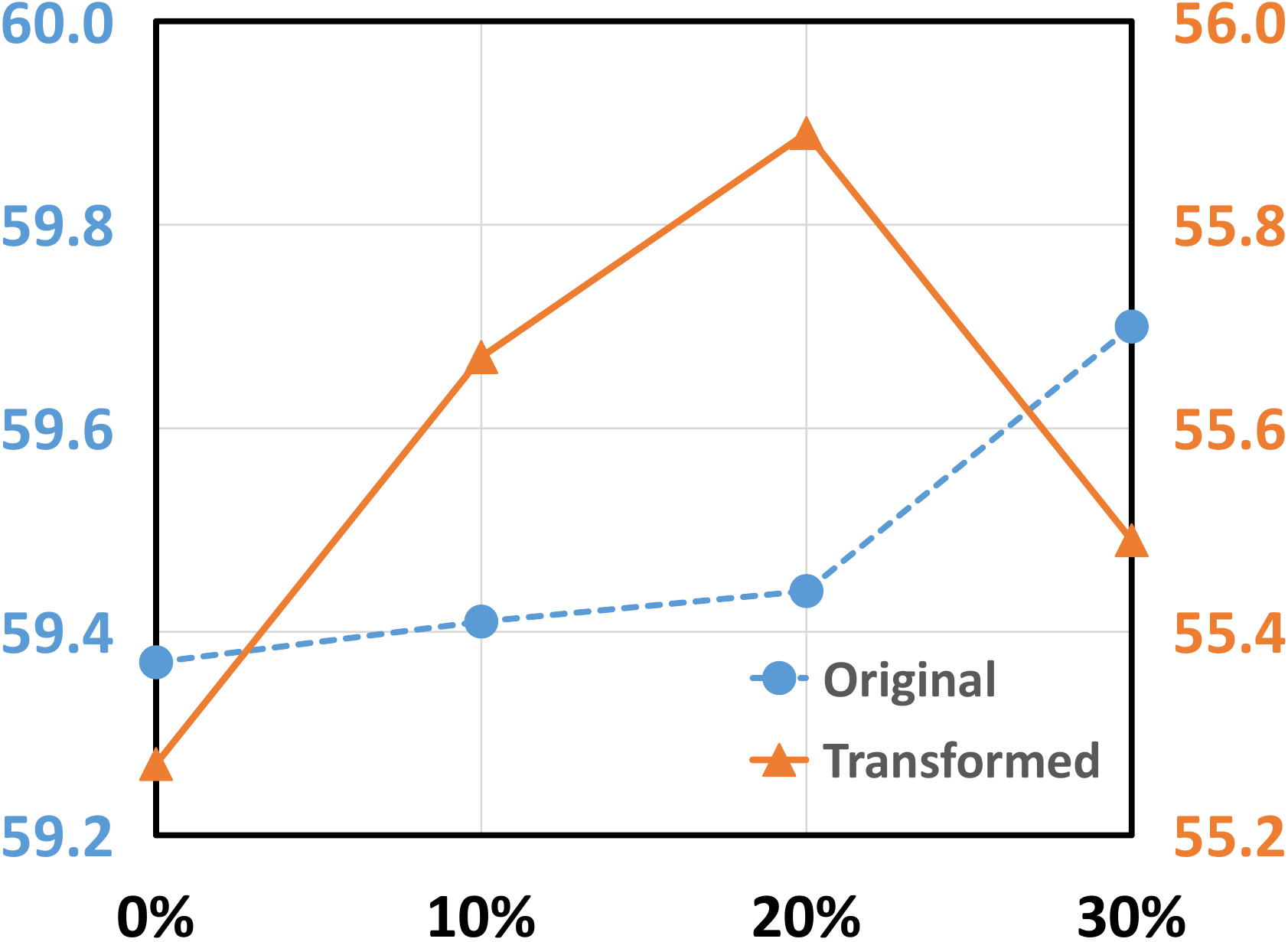}
        \caption{\small $I_{f}$ on defect detection.}
        \end{subfigure}
        \hfill
        \begin{subfigure}[h]{0.235\textwidth}
        \vspace*{0.15cm}
        \centering
        \includegraphics[width=1 \textwidth]{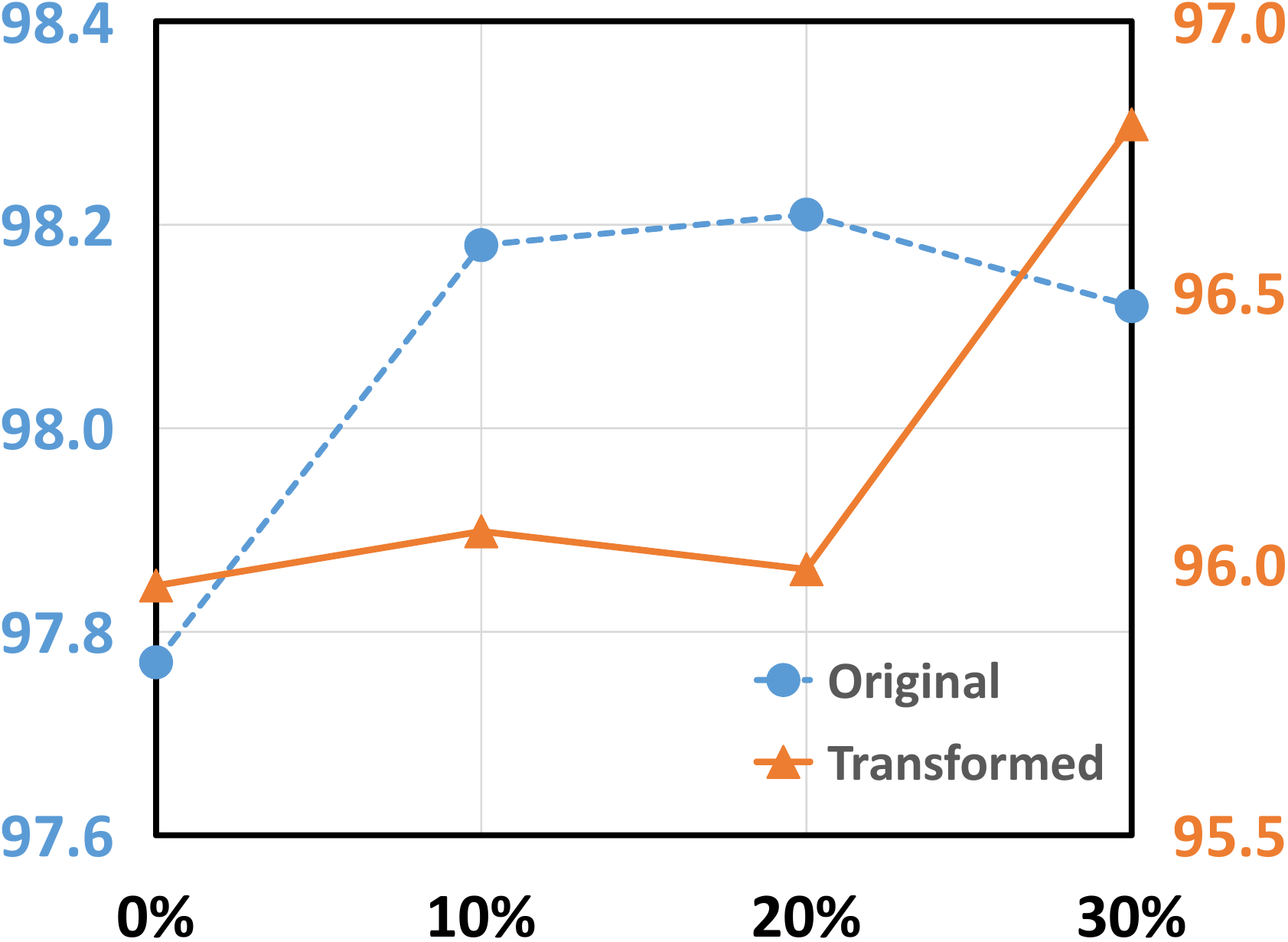}
        \caption{\small $I_{f}$ on code classification.}
        \end{subfigure}
        
         \caption{Parameter analysis on $\alpha$ and $I_{fusion}$. $I_{f}$ is the abbreviation of $I_{fusion}$. The vertical axis means the F1 score, accuracy and accuracy for function naming, defect detection and code classification respectively. The left and right vertical axes indicate
         results on the original and transformed dataset, respectively. 
         }
     \label{fig:parameter}
\end{figure*}

\subsection{RQ3: Ablation Study}\label{sec:ablation}
We further perform ablation studies to verify the effectiveness of two key stages in
\tool, i.e., multi-task learning and counterfactual inference. We select CodeBERT as the basic model, since it is used for evaluation on all the tasks. For function naming, we use the PHP dataset for evaluation.

\textbf{Multi-task learning:} $L_f$ and $L_t$ are introduced
to distinguish the direct causal effect from the \textit{total effect}. From Table~\ref{tab:ablation}, we can observe that models without them suffer from different degree of performance loss on the original test set or transformed test set. Specifically, removing $L_f$ leads to a significant decrease on the transformed test set, {with the decrease rate at}
8.4\%, 5.9\% and 7.9\% for function naming, defect detection and code classification, respectively; while without $L_t$, the framework performance drops consistently on the original test set. 
This indicates that removing $L_f$ prevents the model from fully capturing the misleading information, 
which is harmful to model robustness; while removing $L_t$ makes the model unable to
distinguish \textit{T $\rightarrow$ R} and \textit{K $\rightarrow$ R} well, resulting in poorly exploiting
the beneficial knowledge brought by \textit{K $\rightarrow$ R}. Moreover, removing both $L_f$ and $L_t$ leads to worse performance, e.g., a drop of 3.3\% and 3.6\% on the original and transformed test set of function naming, respectively.

\textbf{Counterfactual inference:} 
We validate the effectiveness of counterfactual inference by setting $\alpha$ in Equ. (\ref{equ:fusion}) to zero. As shown in the last row in Table~\ref{tab:ablation}, without counterfactual inference, the framework's performance decreases on all tasks except a slightly improvement on the original test set for code classification. Specifically, the performance
on the transformed test set drop by 12.7\%, 1.1\%, and 0.4\% on function naming, defect detection and code classification, respectively. The results show that the removal of misleading information by 
counterfactual inference improves the robustness of \tool.



\subsection{RQ4: Parameter Analysis}~\label{subsec:param}
In this section, we analyze how the two key hyper-parameters $\alpha$ and $I_{fusion}$ affect the performance of \tool. 
Due to the page limitation, in figure~\ref{fig:parameter}, we only present the experimental results with CodeBERT on the Go and PHP dataset, TextCNN and ASTNN as basic models for function naming, defect detection, and code classification, respectively. The results on other language and basic models are presented on our GitHub repository.\footnote{{\url{https://github.com/ReliableCoding/CREAM}}}

\textbf{The parameter $\alpha$.} As shown in Figure~\ref{fig:parameter} (a), (b), (c) and (d), the model performance shows similar trend along with the increase of $\alpha$ on the original dataset for all the tasks. \tool' performance first increases and achieves its peak, and then descends obviously with a larger $\alpha$.
The optimal $\alpha$ value is around 0.6 for the tasks. However, for the transformed dataset, the performance increases monotonically as $\alpha$ grows from 0 to 1. 
Recall that $\alpha$ in Equ. (\ref{equ:debias}) is designed to control the degree of eliminating the misleading impact of identifiers, and a larger $\alpha$ will remove more misleading impact of the identifiers. Since the misleading impact in the transformed dataset is more serious than that in the original dataset, a larger $\alpha$ in the transformed dataset is more appreciated.
During the experimentation, to balance the performance of \tool on both original and transformed datasets, we choose the value of $\alpha$ from the set \{0.4, 0.5, 0.6, 0.7, 0.8\}. Specifically, 
we test the model with $\alpha$ set from 0.4 to 0.8, and select the $\alpha$ with best robustness under the condition that the performance on the original dataset is not sacrificed too much. 

\textbf{The parameter $I_{fusion}$.} We study the effect of $I_{fusion}$, as introduced in Section~\ref{subsubsec:framework}, by varying it from 0\% to 30\% of the total training iterations for the tasks. From Figure~\ref{fig:parameter} (e), (f), (g) and (h), 
we can observe that involving $I_{fusion}$ benefits the performance on both original and transformed datasets for all the tasks.
However, for function naming, we also observe that the F1 score when $I_{fusion}$ is set as 30\%
on the original dataset is lower than that when $I_{fusion}$ is set as zero. This may be attributed to that models need more training epochs to well distinguish \textit{T $\rightarrow$ R} and \textit{K $\rightarrow$ R} for the complex generation task.
In this work, we set $I_{fusion}$ as 10\%  of the total training iterations due to the
relatively better results on all tasks.





%% file: section/discussion.tex
\section{Discussion}\label{sec:discuss}

\subsection{Why Counterfactual Inference Helps?}
In this section, we provide another Bayesian perspective to understand how \tool eliminates the misleading impact
of identifiers through counterfactual inference. 
We use $p_t(x,y)$ and $p_s(x,y)$ to denote the data distributions of the training set and test set, respectively. We focus on the common classification tasks~\cite{DBLP:conf/emnlp/Kim14,DBLP:conf/aaai/MouLZWJ16} which normalize classification scores with the softmax function. Based on Bayes' theorem, 
the posterior on the training and test set can be expressed as follows:

\begin{equation}
p_t(y|x) = \frac{p_t(y)p_t(x|y)}{{\textstyle \sum_{y}^{}p_t(y)p_t(x|y)}},
\end{equation}

\begin{equation}
p_s(y|x) = \frac{p_s(y)p_s(x|y)}{{\textstyle \sum_{y}^{}p_s(y)p_s(x|y)}},
\end{equation}
where $p(.)$ is parameterized by a neural network. Our goal is to learn a set of parameters from training set, which is expected to estimate the posterior of test set well. Formally, given a sample $x'$ in test set, we estimate its posterior by:

\begin{equation}
\bar{p}(y|x')  = \frac{p_t(y)p_t(x'|y)}{{\textstyle \sum_{y}^{}p_t(y)p_t(x'|y)}} = \frac{e^{z'}}{\textstyle \sum_{}^{}e^{z'}},
\end{equation}
where $z'$ is the classification score for $x'$. Under random data split, we can assume that the prior on training and test set are the same, i.e. $p_t(y)=p_s(y)$. However, in practice, we cannot ensure that the likelihood on training set $p_t(x|y)$ is the same as that on test set $p_s(x|y)$~\cite{DBLP:journals/csur/GamaZBPB14,DBLP:journals/tkde/LuLDGGZ19}. For example, with respect to defect detection, if functions that contain the identifier ``\textit{ss}'' are all labeled as vulnerable code in the training set, model will learn a higher vulnerable likelihood for the function with ``\textit{ss}'', i.e., $p_t(ss|vulnerable)$ $>$ $p_t(ss|invulnerable)$.
Thus, conventional models that directly predict based on the likelihood learned on the training set may be misled, since the identifier ``\textit{ss}'' is irrelevant to the vulnerability of the function.
Different from conventional models, our framework adaptively estimates and subtracts the misleading impact of identifiers 
from the classification score in the inference stage:

\begin{equation}
\begin{aligned}
\frac{e^{{z'}-f_{z'}}}{\textstyle \sum_{z}^{}e^{{z'}-f_{z'}}} &= \frac{p_t(y)p_t(x'|y){\beta}_y}{{\textstyle \sum_{y}^{}p_t(y)p_t(x'|y){\beta}_y}}\\
\end{aligned}
\end{equation}
where $f_{z'}$ is corresponding to $\alpha * R_{f,k^*,t^*}$ in Equ. (\ref{equ:debias}). Here, the stronger the correlation between the identifiers in $x'$ and $y$, the larger (smaller) the value of $f_{z'}$ (${\beta}_y$) will be.  
This indicates that \tool eliminates the misleading impact of identifiers 
by rectifying 
the incorrect estimation of likelihood learned from the training set.
In this work, we mainly focus on the misleading information 
caused by identifiers, and will explore other potential misleading sources 
in future work.

\subsection{Performance under adversarial attacks}
In this section, we follow previous work~\cite{DBLP:conf/icse/YangSH022,DBLP:journals/corr/abs-2201-07381,DBLP:conf/icse/LiCCZX22} and further evaluate the robustness improvement of \tool by adversarial attack. Specifically, we experiment with CodeBERT on defect detection and code classification since it show the best performance on both tasks. For adversarial attack methods, we select two state-of-the-art black-box methods ALERT~\cite{DBLP:conf/icse/YangSH022} and MHM~\cite{DBLP:conf/aaai/ZhangLLMLJ20} with natural substitution~\cite{DBLP:conf/icse/YangSH022}. We evaluate the robustness of model by the widely-used Attack Success Rates (ASR) metric~\cite{DBLP:conf/cvpr/DongFYPSXZ20,DBLP:conf/icse/YangSH022,DBLP:conf/icse/LiCCZX22}, which measures the fraction of samples that can be attacked among all of the test samples. A higher ASR indicates that a model is more vulnerable to adversarial attack. From Table~\ref{tab:adv_resutls}, we can observe that \tool consistently improves the robustness of CodeBERT on both tasks under both attack approaches. For example, \tool reduces 54.38\% and 55.27\%  possibility of being attacked by ALERT on defect detection and code classification, respectively. This indicates that \tool can also effectively improve the robustness of neural code comprehension models under adversarial attacks. We will experiment with more basic models and adversarial attack methods in our future work.

\begin{table}[t]
\centering
\caption{Comparison results of attack success rates on attacking CodeBERT and CodeBERT+\tool. MHM-NS denotes MHM attack with natural substitution~\cite{DBLP:conf/icse/YangSH022}.}\label{tab:adv_resutls}
\aboverulesep=0ex
\belowrulesep=0ex
\scalebox{1}{
\begin{tabular}{llll}
\toprule
{\textbf{Approach}} & {\textbf{Model}} & \textbf{Defect detection} & \textbf{Code classification} \\
\midrule
\multirow{2}{*}{ALERT} & {CodeBERT}  & 53.97 & 46.43\\
 & { +\tool}  & \textbf{24.62}($\downarrow$54.38\%) & \textbf{20.77}($\downarrow$55.27\%)\\
\hdashline
\multirow{2}{*}{MHM-NS} & {CodeBERT}  & 33.27 & 4.15\\
 & { +\tool}  & \textbf{17.32}($\downarrow$48.21\%) & \textbf{1.21}($\downarrow$70.84\%)\\
\bottomrule
\end{tabular}
}
\end{table}

\subsection{Threats to Validity}

We identify three main treats to validity of our study:

\begin{enumerate}
\item \textbf{The selection of code comprehension tasks.} In this work, we select three popular code comprehension tasks to evaluate \tool , including function naming, defect detection and code classification. Although \tool shows superior performance on these tasks, other tasks such as code search~\cite{DBLP:journals/nn/GuLGWZXL21,DBLP:conf/iwpc/ShuaiX0Y0L20} and code summarization~\cite{10.1145/3522674,DBLP:conf/acl/IyerKCZ16,DBLP:conf/iwpc/HuLXLJ18} are also important and not involved in our experiment. In the future, we will validate \tool on more code comprehension tasks. 

\item \textbf{The selection of basic models.} For each task, we select at least three basic models to validate the effectiveness of \tool. The selected basic models are representative of the corresponding task. To comprehensively evaluate the performance of \tool, more basic models should be considered.
In the future, we will experiment with more basic models to evaluate the generality of \tool.

\item \textbf{The selection of datasets.} For each task, we select one popular dataset for evaluation. However, there are other datasets such as \cite{DBLP:conf/icml/AllamanisPS16} for function naming. In the future, we will conduct experiments on more datasets.

\item \textbf{Comparison to ensemble techniques.} 
\tool aggregates the prediction from three branches with different inputs, which can also be framed as ensemble learning. There is thus a threat that some simple ensemble learning methods can also improve the model's performance. In the future, we will compare \tool with other ensemble techniques to validate the benefits of counterfactual reasoning.


\end{enumerate}

%% file: section/literature.tex
\section{Related work}\label{sec:related}

\subsection{Code Comprehension}
In this section, we focus on deep-learning-based methods on three tasks that are covered in our work including function naming, defect detection and code classification. Besides, the related work on pre-trained models for code are also discussed.

\textbf{Function Naming: } 
Alon et al.~\cite{DBLP:conf/iclr/AlonBLY19} present Code2seq that represents the code snippets by sampling certain paths from the ASTs. Another work proposed by  Z{\"{u}}gner et al.~\cite{DBLP:conf/iclr/ZugnerKCLG21} focuses on multilingual code summarization and proposes to build upon language-agnostic features such as source code and AST-based features. A recent work~\cite{peng2021integrating} propose to encodes tree paths into transformer.

\textbf{Defect Detection: } 
Russell et al.~\cite{DBLP:conf/icmla/RussellKHLHOEM18} empirically evaluate the ML techniques on defect detection and find that TextCNN with an ensemble tree algorithm achieves the best performance. Another work~\cite{DBLP:conf/ndss/LiZXO0WDZ18} proposes the first deep learning-based vulnerability detection system VulDeePecker. Devign~\cite{DBLP:conf/nips/ZhouLSD019} is proposed to learn the various vulnerability characteristics with a composite code property graph and graph neural network.

\textbf{Code classification: }
TBCNN~\cite{DBLP:conf/aaai/MouLZWJ16} is a classical code classification model which captures structural information of the AST with a tree-based convolutional neural network. ASTNN~\cite{DBLP:conf/nips/ZhouLSD019} learns the code representation by splitting the large AST into a sequence of small statement trees. Another recent work~\cite{DBLP:conf/aaai/BuiYJ21} propose to capture the tree structure of code with a capsule network.

\textbf{Pre-trained models for code: } 
Recently, a number of pre-trained models for source code have been proposed~\cite{DBLP:conf/emnlp/FengGTDFGS0LJZ20,DBLP:conf/iclr/GuoRLFT0ZDSFTDC21,DBLP:conf/emnlp/0034WJH21}. CodeBERT~\cite{DBLP:conf/emnlp/FengGTDFGS0LJZ20} is an encoder-only pre-trained model based on Masked language modeling and replaced token detection. GraphCodeBERT~\cite{DBLP:conf/iclr/GuoRLFT0ZDSFTDC21} further leverage code structure information by data flow graph. Another recent work~\cite{DBLP:journals/corr/abs-2201-01549} proposes a sequence-to-sequence pre-trained model with the encoder-decoder architecture.


\subsection{Causal Inference}
Causal inference has attracted increasing attention in fields
including computer vision~\cite{DBLP:conf/iccv/WangZSZ21,DBLP:conf/cvpr/NiuTZL0W21}, natural language processing~\cite{DBLP:conf/aaai/WangC21,DBLP:conf/acl/0003FWMX20} and recommendation~\cite{DBLP:conf/sigir/WangF0ZC21,DBLP:conf/kdd/WeiFCWYH21}. The general purpose of causal inference is to help model pursue causal effect rather than correlation effect. 
Niu et al.~\cite{DBLP:conf/cvpr/NiuTZL0W21} propose a counterfactual framework to remove the language bias in visual question answering. 
In recommendation, \cite{DBLP:conf/sigir/WangF0ZC21} and \cite{DBLP:conf/kdd/WeiFCWYH21} also employ similar methods to eliminate the popularity bias. Some works~\cite{DBLP:conf/sigir/ZhangF0WSL021,DBLP:conf/sigir/YangFJWC21} also consider to build the causal graph from data generation view and 
remove the confounder with back-door adjustment. In text classification, some works~\cite{DBLP:conf/aaai/WangC21,DBLP:conf/acl/Yang0CZSD20} focus on alleviate the spurious correlation by generating counterfactual samples. 
Different from the above studies, we are devoted to extracting and eliminating the misleading impact of identifiers  
in neural code models. To the best of our knowledge, we are the first to introduce the causal inference into neural code comprehension.

%% file: main.bbl
\begin{thebibliography}{10}
\providecommand{\url}[1]{#1}
\csname url@samestyle\endcsname
\providecommand{\newblock}{\relax}
\providecommand{\bibinfo}[2]{#2}
\providecommand{\BIBentrySTDinterwordspacing}{\spaceskip=0pt\relax}
\providecommand{\BIBentryALTinterwordstretchfactor}{4}
\providecommand{\BIBentryALTinterwordspacing}{\spaceskip=\fontdimen2\font plus
\BIBentryALTinterwordstretchfactor\fontdimen3\font minus
  \fontdimen4\font\relax}
\providecommand{\BIBforeignlanguage}[2]{{%
\expandafter\ifx\csname l@#1\endcsname\relax
\typeout{** WARNING: IEEEtran.bst: No hyphenation pattern has been}%
\typeout{** loaded for the language `#1'. Using the pattern for}%
\typeout{** the default language instead.}%
\else
\language=\csname l@#1\endcsname
\fi
#2}}
\providecommand{\BIBdecl}{\relax}
\BIBdecl

\bibitem{DBLP:conf/icse/ZhangWZ0WL19}
J.~Zhang, X.~Wang, H.~Zhang, H.~Sun, K.~Wang, and X.~Liu, ``A novel neural
  source code representation based on abstract syntax tree,'' in
  \emph{Proceedings of the 41st International Conference on Software
  Engineering, {ICSE} 2019}.\hskip 1em plus 0.5em minus 0.4em\relax {IEEE} /
  {ACM}, 2019, pp. 783--794.

\bibitem{DBLP:conf/kbse/LiuLZJ20}
F.~Liu, G.~Li, Y.~Zhao, and Z.~Jin, ``Multi-task learning based pre-trained
  language model for code completion,'' in \emph{35th {IEEE/ACM} International
  Conference on Automated Software Engineering, {ASE} 2020}.\hskip 1em plus
  0.5em minus 0.4em\relax {IEEE}, 2020, pp. 473--485.

\bibitem{DBLP:conf/kbse/WeiLLXJ20}
B.~Wei, Y.~Li, G.~Li, X.~Xia, and Z.~Jin, ``Retrieve and refine: Exemplar-based
  neural comment generation,'' in \emph{35th {IEEE/ACM} International
  Conference on Automated Software Engineering, {ASE} 2020}.\hskip 1em plus
  0.5em minus 0.4em\relax {IEEE}, 2020, pp. 349--360.

\bibitem{DBLP:journals/nn/GuLGWZXL21}
W.~Gu, Z.~Li, C.~Gao, C.~Wang, H.~Zhang, Z.~Xu, and M.~R. Lyu, ``Cradle: Deep
  code retrieval based on semantic dependency learning,'' \emph{Neural
  Networks}, vol. 141, pp. 385--394, 2021.

\bibitem{DBLP:conf/iwpc/ShuaiX0Y0L20}
J.~Shuai, L.~Xu, C.~Liu, M.~Yan, X.~Xia, and Y.~Lei, ``Improving code search
  with co-attentive representation learning,'' in \emph{{ICPC} '20: 28th
  International Conference on Program Comprehension}.\hskip 1em plus 0.5em
  minus 0.4em\relax {ACM}, 2020, pp. 196--207.

\bibitem{DBLP:conf/acl/IyerKCZ16}
S.~Iyer, I.~Konstas, A.~Cheung, and L.~Zettlemoyer, ``Summarizing source code
  using a neural attention model,'' in \emph{Proceedings of the 54th Annual
  Meeting of the Association for Computational Linguistics, {ACL} 2016}.\hskip
  1em plus 0.5em minus 0.4em\relax The Association for Computer Linguistics,
  2016.

\bibitem{DBLP:conf/iwpc/HuLXLJ18}
X.~Hu, G.~Li, X.~Xia, D.~Lo, and Z.~Jin, ``Deep code comment generation,'' in
  \emph{Proceedings of the 26th Conference on Program Comprehension, {ICPC}
  2018, Gothenburg, Sweden, May 27-28, 2018}, F.~Khomh, C.~K. Roy, and
  J.~Siegmund, Eds.\hskip 1em plus 0.5em minus 0.4em\relax {ACM}, 2018, pp.
  200--210.

\bibitem{DBLP:conf/iclr/AlonBLY19}
U.~Alon, S.~Brody, O.~Levy, and E.~Yahav, ``code2seq: Generating sequences from
  structured representations of code,'' in \emph{7th International Conference
  on Learning Representations, {ICLR} 2019}.\hskip 1em plus 0.5em minus
  0.4em\relax OpenReview.net, 2019.

\bibitem{DBLP:conf/iclr/ZugnerKCLG21}
D.~Z{\"{u}}gner, T.~Kirschstein, M.~Catasta, J.~Leskovec, and
  S.~G{\"{u}}nnemann, ``Language-agnostic representation learning of source
  code from structure and context,'' in \emph{9th International Conference on
  Learning Representations, {ICLR} 2021}.\hskip 1em plus 0.5em minus
  0.4em\relax OpenReview.net, 2021.

\bibitem{DBLP:conf/icse/Li0N21}
Y.~Li, S.~Wang, and T.~N. Nguyen, ``A context-based automated approach for
  method name consistency checking and suggestion,'' in \emph{43rd {IEEE/ACM}
  International Conference on Software Engineering, {ICSE} 2021}.\hskip 1em
  plus 0.5em minus 0.4em\relax {IEEE}, 2021, pp. 574--586.

\bibitem{DBLP:conf/nips/ZhouLSD019}
Y.~Zhou, S.~Liu, J.~K. Siow, X.~Du, and Y.~Liu, ``Devign: Effective
  vulnerability identification by learning comprehensive program semantics via
  graph neural networks,'' in \emph{Advances in Neural Information Processing
  Systems 32: Annual Conference on Neural Information Processing Systems 2019,
  NeurIPS 2019}, 2019, pp. 10\,197--10\,207.

\bibitem{DBLP:journals/tosem/ZouZXLJY21}
D.~Zou, Y.~Zhu, S.~Xu, Z.~Li, H.~Jin, and H.~Ye, ``Interpreting deep
  learning-based vulnerability detector predictions based on heuristic
  searching,'' \emph{{ACM} Trans. Softw. Eng. Methodol.}, vol.~30, no.~2, pp.
  23:1--23:31, 2021.

\bibitem{IVDETECT}
Y.~Li, S.~Wang, and T.~N. Nguyen, ``Vulnerability detection with fine-grained
  interpretations,'' in \emph{{ESEC/FSE} '21: 29th {ACM} Joint European
  Software Engineering Conference and Symposium on the Foundations of Software
  Engineering, Athens, Greece, August 23-28, 2021}.\hskip 1em plus 0.5em minus
  0.4em\relax {ACM}, 2021, pp. 292--303.

\bibitem{DBLP:conf/icse/NguyenNNLTP22}
V.~Nguyen, D.~Q. Nguyen, V.~Nguyen, T.~Le, Q.~H. Tran, and D.~Phung, ``Regvd:
  Revisiting graph neural networks for vulnerability detection,'' in \emph{44th
  {IEEE/ACM} International Conference on Software Engineering: Companion
  Proceedings, {ICSE} Companion 2022, Pittsburgh, PA, USA, May 22-24,
  2022}.\hskip 1em plus 0.5em minus 0.4em\relax {ACM/IEEE}, 2022, pp. 178--182.

\bibitem{DBLP:journals/infsof/RabinBWYJA21}
M.~R.~I. Rabin, N.~D.~Q. Bui, K.~Wang, Y.~Yu, L.~Jiang, and M.~A. Alipour, ``On
  the generalizability of neural program models with respect to
  semantic-preserving program transformations,'' \emph{Inf. Softw. Technol.},
  vol. 135, p. 106552, 2021.

\bibitem{DBLP:conf/aaai/ZhangLLMLJ20}
H.~Zhang, Z.~Li, G.~Li, L.~Ma, Y.~Liu, and Z.~Jin, ``Generating adversarial
  examples for holding robustness of source code processing models,'' in
  \emph{The Thirty-Fourth {AAAI} Conference on Artificial Intelligence, {AAAI}
  2020}.\hskip 1em plus 0.5em minus 0.4em\relax {AAAI} Press, 2020, pp.
  1169--1176.

\bibitem{DBLP:journals/pacmpl/Yefet0Y20}
N.~Yefet, U.~Alon, and E.~Yahav, ``Adversarial examples for models of code,''
  \emph{Proc. {ACM} Program. Lang.}, vol.~4, no. {OOPSLA}, pp. 162:1--162:30,
  2020.

\bibitem{zhang2022towards}
H.~Zhang, Z.~Fu, G.~Li, L.~Ma, Z.~Zhao, H.~Yang, Y.~Sun, Y.~Liu, and Z.~Jin,
  ``Towards robustness of deep program processing models--detection, estimation
  and enhancement,'' \emph{ACM Transactions on Software Engineering and
  Methodology}, 2022.

\bibitem{DBLP:journals/corr/abs-2002-03043}
G.~Ramakrishnan, J.~Henkel, Z.~Wang, A.~Albarghouthi, S.~Jha, and T.~W. Reps,
  ``Semantic robustness of models of source code,'' \emph{CoRR}, vol.
  abs/2002.03043, 2020.

\bibitem{DBLP:conf/acl/AhmadCRC20}
W.~U. Ahmad, S.~Chakraborty, B.~Ray, and K.~Chang, ``A transformer-based
  approach for source code summarization,'' in \emph{Proceedings of the 58th
  Annual Meeting of the Association for Computational Linguistics, {ACL}
  2020}.\hskip 1em plus 0.5em minus 0.4em\relax Association for Computational
  Linguistics, 2020, pp. 4998--5007.

\bibitem{DBLP:conf/sigsoft/ChirkovaT21}
N.~Chirkova and S.~Troshin, ``Empirical study of transformers for source
  code,'' in \emph{{ESEC/FSE} '21: 29th {ACM} Joint European Software
  Engineering Conference and Symposium on the Foundations of Software
  Engineering}.\hskip 1em plus 0.5em minus 0.4em\relax {ACM}, 2021, pp.
  703--715.

\bibitem{DBLP:conf/icse/MastropaoloSCNP21}
A.~Mastropaolo, S.~Scalabrino, N.~Cooper, D.~Nader{-}Palacio, D.~Poshyvanyk,
  R.~Oliveto, and G.~Bavota, ``Studying the usage of text-to-text transfer
  transformer to support code-related tasks,'' in \emph{43rd {IEEE/ACM}
  International Conference on Software Engineering, {ICSE} 2021, Madrid, Spain,
  22-30 May 2021}.\hskip 1em plus 0.5em minus 0.4em\relax {IEEE}, 2021, pp.
  336--347.

\bibitem{castelvecchi2016can}
D.~Castelvecchi, ``Can we open the black box of ai?'' \emph{Nature News}, vol.
  538, no. 7623, p.~20, 2016.

\bibitem{schwartz2020green}
R.~Schwartz, J.~Dodge, N.~A. Smith, and O.~Etzioni, ``Green ai,''
  \emph{Communications of the ACM}, vol.~63, no.~12, pp. 54--63, 2020.

\bibitem{zhu2009introduction}
X.~Zhu and A.~B. Goldberg, ``Introduction to semi-supervised learning,''
  \emph{Synthesis lectures on artificial intelligence and machine learning},
  vol.~3, no.~1, pp. 1--130, 2009.

\bibitem{pearl2009causality}
J.~Pearl, \emph{Causality}.\hskip 1em plus 0.5em minus 0.4em\relax Cambridge
  university press, 2009.

\bibitem{mackinnon2007mediation}
D.~P. MacKinnon, A.~J. Fairchild, and M.~S. Fritz, ``Mediation analysis,''
  \emph{Annu. Rev. Psychol.}, vol.~58, pp. 593--614, 2007.

\bibitem{keele2015statistics}
L.~Keele, ``The statistics of causal inference: A view from political
  methodology,'' \emph{Political Analysis}, vol.~23, no.~3, pp. 313--335, 2015.

\bibitem{DBLP:conf/cvpr/NiuTZL0W21}
Y.~Niu, K.~Tang, H.~Zhang, Z.~Lu, X.~Hua, and J.~Wen, ``Counterfactual {VQA:}
  {A} cause-effect look at language bias,'' in \emph{{IEEE} Conference on
  Computer Vision and Pattern Recognition, {CVPR} 2021}.\hskip 1em plus 0.5em
  minus 0.4em\relax Computer Vision Foundation / {IEEE}, 2021, pp.
  12\,700--12\,710.

\bibitem{DBLP:conf/kdd/WeiFCWYH21}
T.~Wei, F.~Feng, J.~Chen, Z.~Wu, J.~Yi, and X.~He, ``Model-agnostic
  counterfactual reasoning for eliminating popularity bias in recommender
  system,'' in \emph{{KDD} '21: The 27th {ACM} {SIGKDD} Conference on Knowledge
  Discovery and Data Mining}.\hskip 1em plus 0.5em minus 0.4em\relax {ACM},
  2021, pp. 1791--1800.

\bibitem{pearl2018book}
J.~Pearl and D.~Mackenzie, \emph{The book of why: the new science of cause and
  effect}.\hskip 1em plus 0.5em minus 0.4em\relax Basic books, 2018.

\bibitem{pearl1998graphs}
J.~Pearl, ``Graphs, causality, and structural equation models,''
  \emph{Sociological Methods \& Research}, vol.~27, no.~2, pp. 226--284, 1998.

\bibitem{glymour2016causal}
M.~Glymour, J.~Pearl, and N.~P. Jewell, \emph{Causal inference in statistics: A
  primer}.\hskip 1em plus 0.5em minus 0.4em\relax John Wiley \& Sons, 2016.

\bibitem{vanderweele2013three}
T.~J. VanderWeele, ``A three-way decomposition of a total effect into direct,
  indirect, and interactive effects,'' \emph{Epidemiology (Cambridge, Mass.)},
  vol.~24, no.~2, p. 224, 2013.

\bibitem{DBLP:journals/corr/HintonVD15}
G.~E. Hinton, O.~Vinyals, and J.~Dean, ``Distilling the knowledge in a neural
  network,'' \emph{CoRR}, vol. abs/1503.02531, 2015.

\bibitem{DBLP:conf/iclr/MenonJRJVK21}
A.~K. Menon, S.~Jayasumana, A.~S. Rawat, H.~Jain, A.~Veit, and S.~Kumar,
  ``Long-tail learning via logit adjustment,'' in \emph{9th International
  Conference on Learning Representations, {ICLR} 2021}.\hskip 1em plus 0.5em
  minus 0.4em\relax OpenReview.net, 2021.

\bibitem{hochreiter1997long}
S.~Hochreiter and J.~Schmidhuber, ``Long short-term memory,'' \emph{Neural
  computation}, vol.~9, no.~8, pp. 1735--1780, 1997.

\bibitem{vaswani2017attention}
A.~Vaswani, N.~Shazeer, N.~Parmar, J.~Uszkoreit, L.~Jones, A.~N. Gomez,
  L.~Kaiser, and I.~Polosukhin, ``Attention is all you need,'' in
  \emph{Advances in Neural Information Processing Systems 30: Annual Conference
  on Neural Information Processing Systems 2017}, 2017, pp. 5998--6008.

\bibitem{DBLP:conf/sigir/WangF0ZC21}
W.~Wang, F.~Feng, X.~He, H.~Zhang, and T.~Chua, ``Clicks can be cheating:
  Counterfactual recommendation for mitigating clickbait issue,'' in
  \emph{{SIGIR} '21: The 44th International {ACM} {SIGIR} Conference on
  Research and Development in Information Retrieval}.\hskip 1em plus 0.5em
  minus 0.4em\relax {ACM}, 2021, pp. 1288--1297.

\bibitem{DBLP:conf/nips/CaoWGAM19}
K.~Cao, C.~Wei, A.~Gaidon, N.~Ar{\'{e}}chiga, and T.~Ma, ``Learning imbalanced
  datasets with label-distribution-aware margin loss,'' in \emph{Advances in
  Neural Information Processing Systems 32: Annual Conference on Neural
  Information Processing Systems 2019, NeurIPS 2019}, 2019, pp. 1565--1576.

\bibitem{DBLP:conf/icse/NguyenPLN20}
S.~Nguyen, H.~Phan, T.~Le, and T.~N. Nguyen, ``Suggesting natural method names
  to check name consistencies,'' in \emph{{ICSE} '20: 42nd International
  Conference on Software Engineering}.\hskip 1em plus 0.5em minus 0.4em\relax
  {ACM}, 2020, pp. 1372--1384.

\bibitem{DBLP:journals/corr/abs-2201-10705}
F.~Liu, G.~Li, Z.~Fu, S.~Lu, Y.~Hao, and Z.~Jin, ``Learning to recommend method
  names with global context,'' \emph{CoRR}, vol. abs/2201.10705, 2022.

\bibitem{bishop2006pattern}
C.~M. Bishop and N.~M. Nasrabadi, \emph{Pattern recognition and machine
  learning}.\hskip 1em plus 0.5em minus 0.4em\relax Springer, 2006, vol.~4,
  no.~4.

\bibitem{DBLP:conf/aaai/MouLZWJ16}
L.~Mou, G.~Li, L.~Zhang, T.~Wang, and Z.~Jin, ``Convolutional neural networks
  over tree structures for programming language processing,'' in
  \emph{Proceedings of the Thirtieth {AAAI} Conference on Artificial
  Intelligence}.\hskip 1em plus 0.5em minus 0.4em\relax {AAAI} Press, 2016, pp.
  1287--1293.

\bibitem{DBLP:conf/emnlp/FengGTDFGS0LJZ20}
Z.~Feng, D.~Guo, D.~Tang, N.~Duan, X.~Feng, M.~Gong, L.~Shou, B.~Qin, T.~Liu,
  D.~Jiang, and M.~Zhou, ``Codebert: {A} pre-trained model for programming and
  natural languages,'' in \emph{Findings of the Association for Computational
  Linguistics: {EMNLP} 2020}, ser. Findings of {ACL}, vol. {EMNLP} 2020.\hskip
  1em plus 0.5em minus 0.4em\relax Association for Computational Linguistics,
  2020, pp. 1536--1547.

\bibitem{DBLP:journals/corr/abs-2102-04664}
S.~Lu, D.~Guo, S.~Ren, J.~Huang, A.~Svyatkovskiy, A.~Blanco, C.~B. Clement,
  D.~Drain, D.~Jiang, D.~Tang, G.~Li, L.~Zhou, L.~Shou, L.~Zhou, M.~Tufano,
  M.~Gong, M.~Zhou, N.~Duan, N.~Sundaresan, S.~K. Deng, S.~Fu, and S.~Liu,
  ``Codexglue: {A} machine learning benchmark dataset for code understanding
  and generation,'' \emph{CoRR}, vol. abs/2102.04664, 2021.

\bibitem{DBLP:conf/emnlp/Kim14}
Y.~Kim, ``Convolutional neural networks for sentence classification,'' in
  \emph{Proceedings of the 2014 Conference on Empirical Methods in Natural
  Language Processing, {EMNLP} 2014}.\hskip 1em plus 0.5em minus 0.4em\relax
  {ACL}, 2014, pp. 1746--1751.

\bibitem{DBLP:conf/naacl/YangYDHSH16}
Z.~Yang, D.~Yang, C.~Dyer, X.~He, A.~J. Smola, and E.~H. Hovy, ``Hierarchical
  attention networks for document classification,'' in \emph{{NAACL} {HLT}
  2016, The 2016 Conference of the North American Chapter of the Association
  for Computational Linguistics: Human Language Technologies}.\hskip 1em plus
  0.5em minus 0.4em\relax The Association for Computational Linguistics, 2016,
  pp. 1480--1489.

\bibitem{DBLP:conf/ndss/LiZXO0WDZ18}
Z.~Li, D.~Zou, S.~Xu, X.~Ou, H.~Jin, S.~Wang, Z.~Deng, and Y.~Zhong,
  ``Vuldeepecker: {A} deep learning-based system for vulnerability detection,''
  in \emph{25th Annual Network and Distributed System Security Symposium,
  {NDSS} 2018}.\hskip 1em plus 0.5em minus 0.4em\relax The Internet Society,
  2018.

\bibitem{DBLP:conf/icmla/RussellKHLHOEM18}
R.~L. Russell, L.~Y. Kim, L.~H. Hamilton, T.~Lazovich, J.~Harer, O.~Ozdemir,
  P.~M. Ellingwood, and M.~W. McConley, ``Automated vulnerability detection in
  source code using deep representation learning,'' in \emph{17th {IEEE}
  International Conference on Machine Learning and Applications, {ICMLA}
  2018}.\hskip 1em plus 0.5em minus 0.4em\relax {IEEE}, 2018, pp. 757--762.

\bibitem{DBLP:journals/corr/abs-1909-09436}
H.~Husain, H.~Wu, T.~Gazit, M.~Allamanis, and M.~Brockschmidt, ``Codesearchnet
  challenge: Evaluating the state of semantic code search,'' \emph{CoRR}, vol.
  abs/1909.09436, 2019.

\bibitem{DBLP:conf/emnlp/0034WJH21}
Y.~Wang, W.~Wang, S.~R. Joty, and S.~C.~H. Hoi, ``Codet5: Identifier-aware
  unified pre-trained encoder-decoder models for code understanding and
  generation,'' in \emph{Proceedings of the 2021 Conference on Empirical
  Methods in Natural Language Processing, {EMNLP} 2021}.\hskip 1em plus 0.5em
  minus 0.4em\relax Association for Computational Linguistics, 2021, pp.
  8696--8708.

\bibitem{DBLP:conf/sigir/BuiYJ21}
N.~D.~Q. Bui, Y.~Yu, and L.~Jiang, ``Self-supervised contrastive learning for
  code retrieval and summarization via semantic-preserving transformations,''
  in \emph{{SIGIR} '21: The 44th International {ACM} {SIGIR} Conference on
  Research and Development in Information Retrieval}.\hskip 1em plus 0.5em
  minus 0.4em\relax {ACM}, 2021, pp. 511--521.

\bibitem{DBLP:journals/corr/abs-2111-10793}
W.~Zhang, S.~Guo, H.~Zhang, Y.~Sui, Y.~Xue, and Y.~Xu, ``Challenging machine
  learning-based clone detectors via semantic-preserving code
  transformations,'' \emph{CoRR}, vol. abs/2111.10793, 2021.

\bibitem{DBLP:conf/issta/ZengTZLZZ22}
Z.~Zeng, H.~Tan, H.~Zhang, J.~Li, Y.~Zhang, and L.~Zhang, ``An extensive study
  on pre-trained models for program understanding and generation,'' in
  \emph{{ISSTA} '22: 31st {ACM} {SIGSOFT} International Symposium on Software
  Testing and Analysis, Virtual Event, South Korea, July 18 - 22, 2022}, S.~Ryu
  and Y.~Smaragdakis, Eds.\hskip 1em plus 0.5em minus 0.4em\relax {ACM}, 2022,
  pp. 39--51.

\bibitem{reveal}
S.~Chakraborty, R.~Krishna, Y.~Ding, and B.~Ray, ``Deep learning based
  vulnerability detection: Are we there yet,'' \emph{IEEE Transactions on
  Software Engineering}, 2021.

\bibitem{DBLP:journals/csur/WangYKN20}
Y.~Wang, Q.~Yao, J.~T. Kwok, and L.~M. Ni, ``Generalizing from a few examples:
  {A} survey on few-shot learning,'' \emph{{ACM} Comput. Surv.}, vol.~53,
  no.~3, pp. 63:1--63:34, 2020.

\bibitem{DBLP:conf/iclr/ZhangBHRV17}
C.~Zhang, S.~Bengio, M.~Hardt, B.~Recht, and O.~Vinyals, ``Understanding deep
  learning requires rethinking generalization,'' in \emph{5th International
  Conference on Learning Representations, {ICLR} 2017, Toulon, France, April
  24-26, 2017, Conference Track Proceedings}.\hskip 1em plus 0.5em minus
  0.4em\relax OpenReview.net, 2017.

\bibitem{DBLP:journals/csur/GamaZBPB14}
J.~Gama, I.~Zliobaite, A.~Bifet, M.~Pechenizkiy, and A.~Bouchachia, ``A survey
  on concept drift adaptation,'' \emph{{ACM} Comput. Surv.}, vol.~46, no.~4,
  pp. 44:1--44:37, 2014.

\bibitem{DBLP:journals/tkde/LuLDGGZ19}
J.~Lu, A.~Liu, F.~Dong, F.~Gu, J.~Gama, and G.~Zhang, ``Learning under concept
  drift: {A} review,'' \emph{{IEEE} Trans. Knowl. Data Eng.}, vol.~31, no.~12,
  pp. 2346--2363, 2019.

\bibitem{DBLP:conf/icse/YangSH022}
Z.~Yang, J.~Shi, J.~He, and D.~Lo, ``Natural attack for pre-trained models of
  code,'' in \emph{44th {IEEE/ACM} 44th International Conference on Software
  Engineering, {ICSE} 2022, Pittsburgh, PA, USA, May 25-27, 2022}.\hskip 1em
  plus 0.5em minus 0.4em\relax {ACM}, 2022, pp. 1482--1493.

\bibitem{DBLP:journals/corr/abs-2201-07381}
Z.~Li, Y.~Li, T.~Li, M.~Du, B.~Wu, Y.~Cao, X.~Xie, Y.~Li, and Y.~Liu,
  ``Unveiling project-specific bias in neural code models,'' \emph{CoRR}, vol.
  abs/2201.07381, 2022.

\bibitem{DBLP:conf/icse/LiCCZX22}
Z.~Li, Q.~G. Chen, C.~Chen, Y.~Zou, and S.~Xu, ``Ropgen: Towards robust code
  authorship attribution via automatic coding style transformation,'' in
  \emph{44th {IEEE/ACM} 44th International Conference on Software Engineering,
  {ICSE} 2022, Pittsburgh, PA, USA, May 25-27, 2022}.\hskip 1em plus 0.5em
  minus 0.4em\relax {ACM}, 2022, pp. 1906--1918.

\bibitem{DBLP:conf/cvpr/DongFYPSXZ20}
Y.~Dong, Q.~Fu, X.~Yang, T.~Pang, H.~Su, Z.~Xiao, and J.~Zhu, ``Benchmarking
  adversarial robustness on image classification,'' in \emph{2020 {IEEE/CVF}
  Conference on Computer Vision and Pattern Recognition, {CVPR} 2020, Seattle,
  WA, USA, June 13-19, 2020}.\hskip 1em plus 0.5em minus 0.4em\relax Computer
  Vision Foundation / {IEEE}, 2020, pp. 318--328.

\bibitem{10.1145/3522674}
S.~Gao, C.~Gao, Y.~He, J.~Zeng, L.~Y. Nie, X.~Xia, and M.~R. Lyu, ``Code
  structure guided transformer for source code summarization,'' \emph{ACM
  Trans. Softw. Eng. Methodol.}, 2022.

\bibitem{DBLP:conf/icml/AllamanisPS16}
M.~Allamanis, H.~Peng, and C.~Sutton, ``A convolutional attention network for
  extreme summarization of source code,'' in \emph{Proceedings of the 33nd
  International Conference on Machine Learning, {ICML} 2016}, ser. {JMLR}
  Workshop and Conference Proceedings, vol.~48.\hskip 1em plus 0.5em minus
  0.4em\relax JMLR.org, 2016, pp. 2091--2100.

\bibitem{peng2021integrating}
H.~Peng, G.~Li, W.~Wang, Y.~Zhao, and Z.~Jin, ``Integrating tree path in
  transformer for code representation,'' \emph{Advances in Neural Information
  Processing Systems}, vol.~34, 2021.

\bibitem{DBLP:conf/aaai/BuiYJ21}
N.~D.~Q. Bui, Y.~Yu, and L.~Jiang, ``Treecaps: Tree-based capsule networks for
  source code processing,'' in \emph{Thirty-Fifth {AAAI} Conference on
  Artificial Intelligence, {AAAI} 2021}.\hskip 1em plus 0.5em minus 0.4em\relax
  {AAAI} Press, 2021, pp. 30--38.

\bibitem{DBLP:conf/iclr/GuoRLFT0ZDSFTDC21}
D.~Guo, S.~Ren, S.~Lu, Z.~Feng, D.~Tang, S.~Liu, L.~Zhou, N.~Duan,
  A.~Svyatkovskiy, S.~Fu, M.~Tufano, S.~K. Deng, C.~B. Clement, D.~Drain,
  N.~Sundaresan, J.~Yin, D.~Jiang, and M.~Zhou, ``Graphcodebert: Pre-training
  code representations with data flow,'' in \emph{9th International Conference
  on Learning Representations, {ICLR} 2021}.\hskip 1em plus 0.5em minus
  0.4em\relax OpenReview.net, 2021.

\bibitem{DBLP:journals/corr/abs-2201-01549}
C.~Niu, C.~Li, V.~Ng, J.~Ge, L.~Huang, and B.~Luo, ``Spt-code:
  Sequence-to-sequence pre-training for learning source code representations,''
  \emph{CoRR}, vol. abs/2201.01549, 2022.

\bibitem{DBLP:conf/iccv/WangZSZ21}
T.~Wang, C.~Zhou, Q.~Sun, and H.~Zhang, ``Causal attention for unbiased visual
  recognition,'' in \emph{2021 {IEEE/CVF} International Conference on Computer
  Vision, {ICCV} 2021}.\hskip 1em plus 0.5em minus 0.4em\relax {IEEE}, 2021,
  pp. 3071--3080.

\bibitem{DBLP:conf/aaai/WangC21}
Z.~Wang and A.~Culotta, ``Robustness to spurious correlations in text
  classification via automatically generated counterfactuals,'' in
  \emph{Thirty-Fifth {AAAI} Conference on Artificial Intelligence, {AAAI}
  2021}.\hskip 1em plus 0.5em minus 0.4em\relax {AAAI} Press, 2021, pp.
  14\,024--14\,031.

\bibitem{DBLP:conf/acl/0003FWMX20}
C.~Qian, F.~Feng, L.~Wen, C.~Ma, and P.~Xie, ``Counterfactual inference for
  text classification debiasing,'' in \emph{Proceedings of the 59th Annual
  Meeting of the Association for Computational Linguistics and the 11th
  International Joint Conference on Natural Language Processing, {ACL/IJCNLP}
  2021}.\hskip 1em plus 0.5em minus 0.4em\relax Association for Computational
  Linguistics, 2021, pp. 5434--5445.

\bibitem{DBLP:conf/sigir/ZhangF0WSL021}
Y.~Zhang, F.~Feng, X.~He, T.~Wei, C.~Song, G.~Ling, and Y.~Zhang, ``Causal
  intervention for leveraging popularity bias in recommendation,'' in
  \emph{{SIGIR} '21: The 44th International {ACM} {SIGIR} Conference on
  Research and Development in Information Retrieval}.\hskip 1em plus 0.5em
  minus 0.4em\relax {ACM}, 2021, pp. 11--20.

\bibitem{DBLP:conf/sigir/YangFJWC21}
X.~Yang, F.~Feng, W.~Ji, M.~Wang, and T.~Chua, ``Deconfounded video moment
  retrieval with causal intervention,'' in \emph{{SIGIR} '21: The 44th
  International {ACM} {SIGIR} Conference on Research and Development in
  Information Retrieval}.\hskip 1em plus 0.5em minus 0.4em\relax {ACM}, 2021,
  pp. 1--10.

\bibitem{DBLP:conf/acl/Yang0CZSD20}
L.~Yang, J.~Li, P.~Cunningham, Y.~Zhang, B.~Smyth, and R.~Dong, ``Exploring the
  efficacy of automatically generated counterfactuals for sentiment analysis,''
  in \emph{Proceedings of the 59th Annual Meeting of the Association for
  Computational Linguistics and the 11th International Joint Conference on
  Natural Language Processing, {ACL/IJCNLP} 2021}.\hskip 1em plus 0.5em minus
  0.4em\relax Association for Computational Linguistics, 2021, pp. 306--316.

\end{thebibliography}
